\documentclass[useAMS,usenatbib]{mnras}
\usepackage{graphicx}
\usepackage{url}
\usepackage[T1]{fontenc}
\usepackage{amsmath}
\usepackage{amssymb}

\def\btheta{\mbox{\boldmath$\theta$}}

\title[Studies of Two PCE Binary Stars]
      {Studies of Two Post Common Envelope Binary Stars}

\author[C. Koen and A. Kniazev]{%
    C.~Koen$^{1}$\thanks{E-mail: ckoen@uwc.ac.za}
    and A.~Kniazev$^{2,3,4}$\\
    $^{1}$Department of Statistics, University of the Western Cape,
          Private Bag X17, Bellville 7535, South Africa\\
    $^{2}$South African Astronomical Observatory, PO Box 9,
          Observatory 7935, South Africa\\
    $^{3}$Southern African Large Telescope, PO Box 9,
          Observatory 7935, South Africa\\
    $^{4}$Sternberg Astronomical Institute, Lomonosov Moscow State University,
          Moscow, Russia
}

\begin{document}

\date{Accepted 2026. Received 2026; in original form 2026}
\pagerange{\pageref{firstpage}--\pageref{lastpage}}
\pubyear{2026}

\maketitle
\label{firstpage}

\begin{abstract}
New time series photometry of the short period ($P<0.2$~d)
binary stars
ZTF~J032906.36+070408.3 (ZTF~0329) and ATO~J090.7238+00.5996
(ATL~0602) is presented. The primary (hotter) stars in both systems
are white dwarfs, while the cooler stars are M dwarfs. Archival photometry
and new and archival spectra are used to derive the temperatures of the
components in the two binaries. Both are single-lined binaries. Velocity
curves are fitted to radial velocity measurements of the red dwarfs and
used to place lower limits on the masses of the white dwarfs.
New and archival time series photometry show evolution in the shapes
of the light curves of both stars.
Spectra of ZTF~0329 show an excess of radiation in the $R_C$ band which
is ascribed to electron cyclotron radiation, suggesting that the star is
a low accretion rate polar.
Satisfactory model fits to the multi-filter light curves of ATL~0602
require the presence of cool spots on the red component of the binary.
The latter star exhibits a variety of velocities associated with Balmer
and Ca~II emission lines within individual spectra.
\end{abstract}

\begin{keywords}
stars: binaries: close -- stars: variables: general --
stars: individual: ZTF~J032906.36+070408.3, ATO~J090.7238+00.5996
\end{keywords}

\section{Introduction}
\label{sec:intro}

Variability in ZTF~J032906.36+070408.3 (hereafter ZTF~0329) was discovered
by \citet{Chen2020}, who classified the star as a contact binary with an
$r$-band period of 0.1437522~d.
ATO~J090.7238+00.5996 (2000 sexagesimal coordinates 06:02:53.73, +00:35:58.3;
hereafter referred to as ATL~0602) was first identified as a variable star
by \citet{Heinze2018} who described the lightcurve as a ``modulated sinusoid''
with period 0.196423~d. The star is classified as an RS~CVn spotted rotator in
the Zwicky Transient Facility (ZTF) catalogue of periodic variables
\citep{Chen2020}.

Both stars were chosen for further observation as part of a program to obtain
well-defined light curves of short period main sequence binaries
\citep{Koen2022a,Koen2022b}. Aside from the short periods, the candidate
selection requires {\it Gaia} colour $B_p-R_p>1.84$, which corresponds to
main sequence spectral types M0 and later. However, experience has shown that
binaries comprised of red dwarfs (RDs) with cool white dwarf (WD) companions
may also have red {\it Gaia} colour indices \citep[e.g.][]{Koen2022b}. This
happens to be the case for both stars in the present study.

Given that the WD components has evolved beyond their main sequence lifetimes,
and given the very short periods of the systems, both binaries are evidently in
post common envelope binary (PCEB) phases of evolution. The originally more
massive component in each system has evolved to overfill its Roche equipotential
surface. The consequent mass overflow lead to a common envelope for the two
stars. Loss of the envelope left the current configuration of a close binary
consisting of a WD with a cool main sequence companion. Further evolution of
the systems will be driven by angular momentum loss, which will lead to a
further reduction in the period. More detailed discussions can be found in
\citet[e.g.][]{Ritter2012,Liu2023}.
PCEBs are of interest for a number of reasons, perhaps the most evident being
as sources of information about the short-lived common envelope phase of
evolution.

The material presented below suggests that ZTF~0329 is a low accretion rate
polar \citep[e.g.][]{Schwope2002,Schwope2025}. In this system the WD has a
strong magnetic field and accretes mass from the RD, most likely via a stellar
wind. Both photometric and spectroscopic evidence supporting this thesis is
discussed.

Emission lines from ATL~0602 appear to originate from different locations in
the binary system, even though evidently associated with the RD component.
The star is probably in a semi-detached state, likely to evolve into a
cataclysmic variable, in which mass flows from the RD to the WD through the
inner Lagrangian point.

\section{Observations}
\label{sec:obs}

\subsection{Photometry}
\label{sec:phot}

Time series photometry of the two stars was acquired with CCD cameras attached
to the South African Astronomical Observatory 1.9~m and 1.0~m telescopes
situated near Sutherland, South Africa -- see the log in Table~\ref{tab:photlog}.
In cases where more than one filter is listed for a given run, measurements were
cycled through the filters.

All the observations were made with SHOC CCD cameras \citep{Coppejans2013}.
Measurements were made through standard Johnson-Cousins $VR_CI_C$ filters,
and for two runs on ATL~0602, a Johnson $B$ filter.
(For convenience, the subscripts on the names of the $R_C$ and $I_C$ filters
will be omitted in the remainder of the paper.)
Observing conditions were variable, due to changing sky transparency and
brightness, and changes in seeing (in the range 1--3 arcseconds).
Exposure times and the prebinning configuration (either $2 \times 2$ or
$4 \times 4$) were tailored accordingly. All photometry was differentially
corrected, using constant comparison stars in the field of view.

\begin{figure*}
\centering
\includegraphics[width=\textwidth]{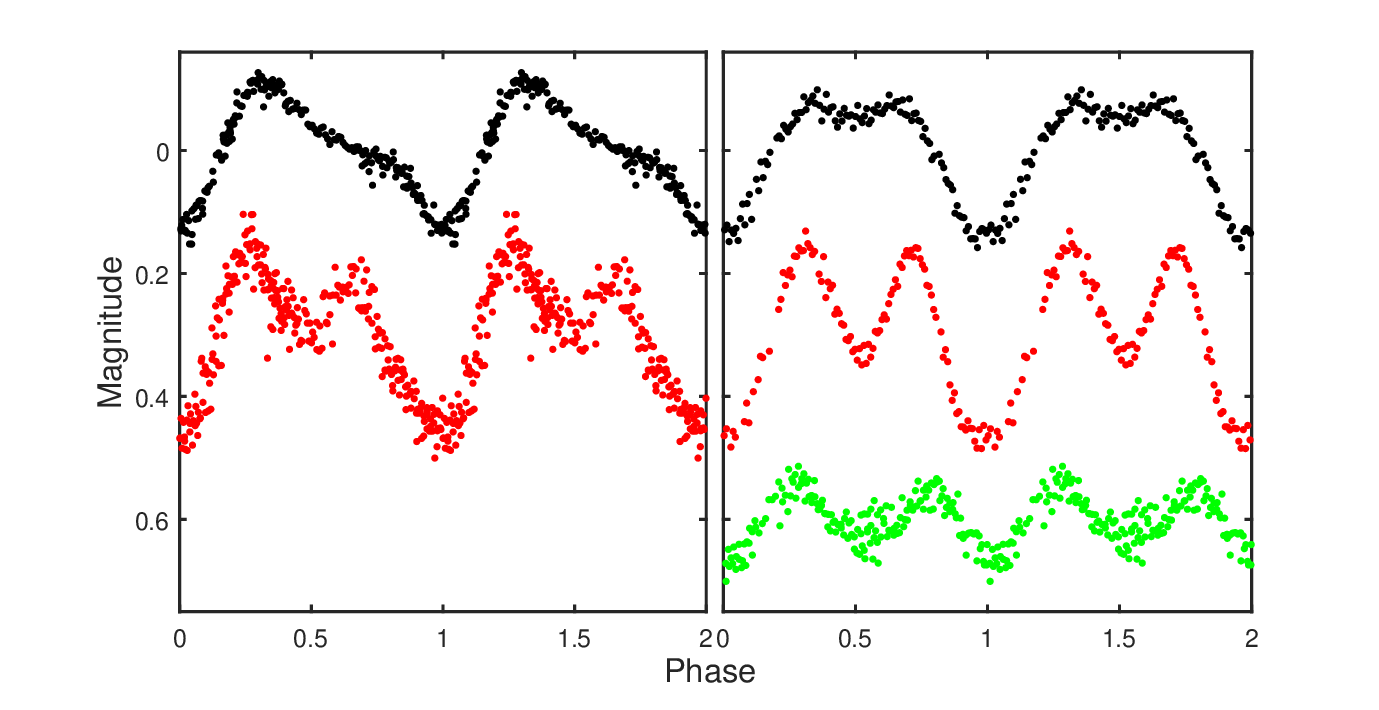}
\caption{Phase-folded light curves of ZTF~0329 obtained at SAAO.
         Left panel: 2022--2023 observations. Right panel: 2025 observations.
         From top to bottom $I$, $R$ and $V$ (right panel only).
         Zeropoints are arbitrary.}
\label{fig:ztf0329lc}
\end{figure*}

\begin{figure*}
\centering
\includegraphics[width=\textwidth]{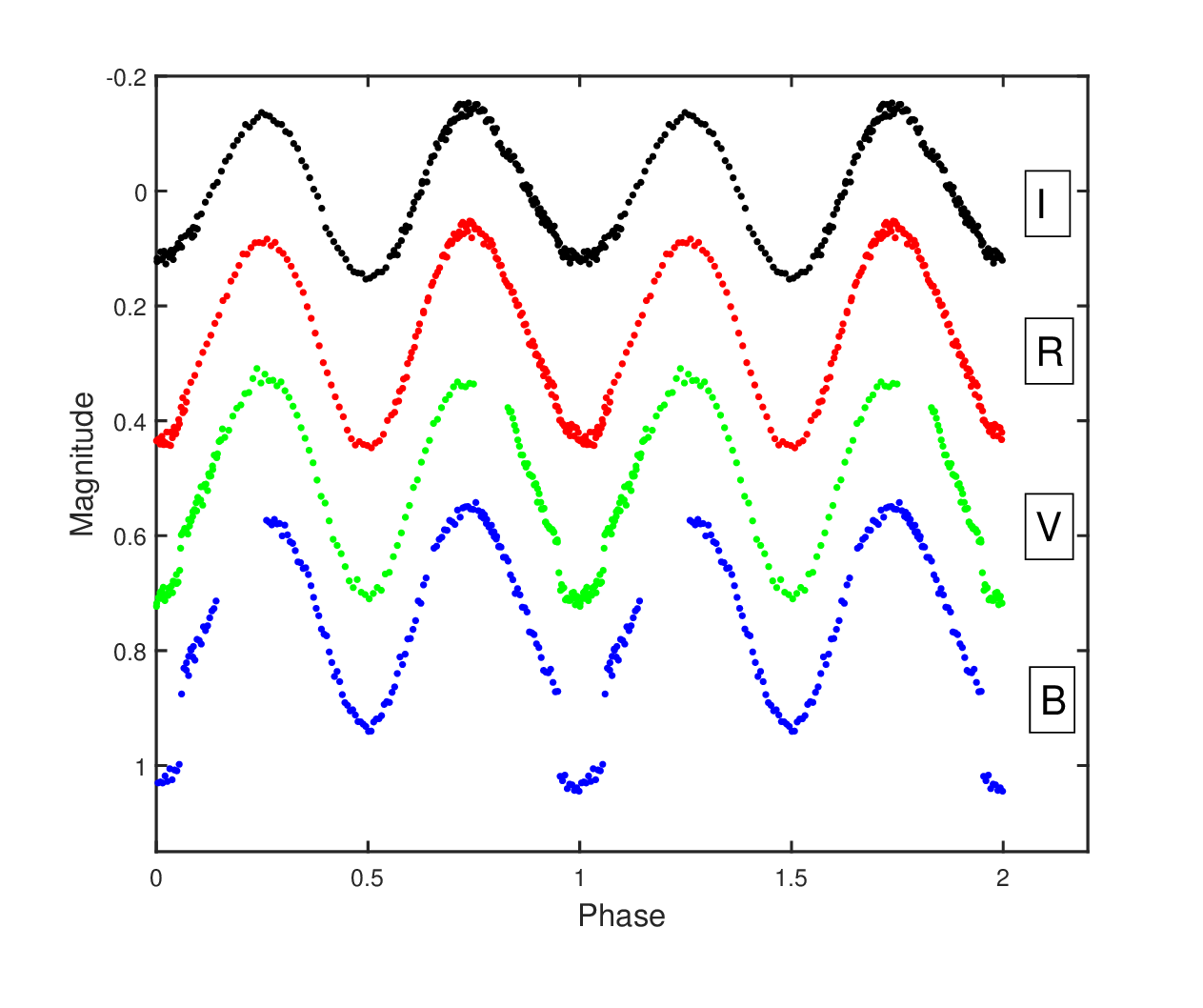}
\caption{Phase-folded light curves of ATL~0602 obtained at SAAO.
         Zeropoints are arbitrary.}
\label{fig:atl0602lc}
\end{figure*}

Reductions were performed concurrently with the observing, using an automated
version of DOPHOT \citep{Schechter1993}. Point spread function magnitudes were
found to have less scatter than those calculated from aperture photometry.
Phase-folded light curves are plotted in Figs~\ref{fig:ztf0329lc}
and~\ref{fig:atl0602lc}; these will be discussed in detail below.

\begin{table*}
\centering
\caption{The photometric observing log. The last column contains the number
         of observations through each of the filters.}
\label{tab:photlog}
\begin{tabular}{cccccl}
\hline\hline
Telescope & Year-Month & Starting time          & Filters & Run length & $N$ \\
(m)       &            & (HJD~2450000+)         &         & (h)        &     \\
\hline
\multicolumn{6}{c}{ZTF~0329}\\
1.9 & 2022-09 & 9831.56033  & $RI$   & 2.6 & 54, 56 \\
1.9 & 2022-09 & 9832.58712  & $RI$   & 1.9 & 26, 26 \\
1.0 & 2022-12 & 9935.29978  & $R$    & 3.6 & 88     \\
1.0 & 2022-12 & 9936.28213  & $I$    & 3.6 & 117    \\
1.0 & 2023-12 & 10299.29003 & $R$    & 3.7 & 108    \\
1.9 & 2025-12 & 11029.29578 & $V$    & 3.7 & 84     \\
1.9 & 2025-12 & 11034.28304 & $V$    & 3.1 & 89     \\
1.9 & 2025-12 & 11035.28768 & $RI$   & 3.4 & 104, 113 \\
\hline
\multicolumn{6}{c}{ATL~0602}\\
1.0 & 2020-12 & 9206.46447  & $RI$   & 2.5 & 107, 112 \\
1.9 & 2025-12 & 11039.34776 & $VRI$  & 4.1 & 90, 90, 92 \\
1.9 & 2025-12 & 11039.52222 & $B$    & 1.5 & 58  \\
1.9 & 2025-12 & 11040.36723 & $B$    & 3.5 & 103 \\
1.9 & 2025-12 & 11040.51634 & $V$    & 1.5 & 85  \\
\hline
\end{tabular}
\end{table*}

\subsection{SALT long-slit spectroscopy}
\label{sec:salt}

Spectral observations were carried out using the South African Large
Telescope \citep[SALT;][]{Buckley2006,ODonoghue2006} using the Robert
Stoby spectrograph \citep[RSS;][]{Burgh2003,Kobulnicky2003} in long-slit
mode. The position angle for the observations was always set to zero. Since
SALT is equipped with an atmospheric dispersion corrector (ADC), this does not
lead to problems with atmospheric dispersion. The names of the observed stars,
dates, exposure times, the grism used, visibility conditions during the
observations and other parameters of our observations are presented in
Table~\ref{tab:saltlog}. Reference spectra were observed immediately after the
scientific exposures. For relative flux calibration, spectrophotometric standard
stars were observed with the same spectral settings during the nearest twilight
as part of the standard SALT spectral data calibrations. It should be noted here
that absolute flux calibration using SALT is not possible, as the telescope's
unfilled entrance pupil shifts during observations; however, the relative
spectral distributions are very accurate, as the telescope optics are stable.
Initial data processing was carried out using the standard SALT pipeline
\citep{Kotze2025}, and subsequent spectral processing was performed using the
method described in \citet{Kniazev2022}.

The spectral range 5085--6040~\AA\ was chosen for SALT observations to encompass
the $\mathrm{Mg}\,b$ triplet (wavelengths 5167.32, 5172.68, 5183.60~\AA),
which serves as a metallicity indicator and is well-resolved at the spectral
resolution of our observations ($R \approx 4200$, see Table~\ref{tab:saltlog}).
Two spectra over the range 6033--6850~\AA\ were obtained to monitor
H$\alpha$ emission.

\begin{table*}
\centering
\caption{The SALT spectroscopic observing log. The last two spectra of ZTF~0329
         are referred to as SALT1 and SALT2 in the rest of the paper.}
\label{tab:saltlog}
\begin{tabular}{lccrccc}
\hline\hline
Date              & Exposure       & Grating & $R$  & Slit width             & Seeing   & Spectral range \\
                  & (sec)          &         &      & (arcsec\,pixel$^{-1}$) & (arcsec) & (\AA)          \\
\hline
\multicolumn{7}{c}{ZTF~0329} \\
26 December 2024  &  1610          & PG2300  & 4200 & 1.5  & 2.0 & 6033--6850 \\
27 December 2024  &  1920          & PG2300  & 4200 & 1.5  & 2.0 & 6033--6850 \\
 3 January   2026 &  2100          & PG0700  &  740 & 1.5  & 1.8 & 3600--7450 \\
 6 January   2026 & $900\times 2$  & PG0700  &  740 & 1.5  & 1.9 & 3600--7450 \\
\hline
\multicolumn{7}{c}{ATL~0602} \\
24 December 2023  &  900           & PG2300  & 3750 & 1.25 & 2.5 & 5085--6040 \\
21 January  2024  &  600           & PG2300  & 3750 & 1.25 & 1.8 & 5085--6040 \\
22 February 2024  &  600           & PG2300  & 3750 & 1.25 & 1.9 & 5085--6040 \\
26 March    2024  &  600           & PG2300  & 3750 & 1.25 & 2.3 & 5085--6040 \\
15 November 2024  & 1620           & PG2300  & 3750 & 1.25 & 2.0 & 5085--6040 \\
24 November 2024  & 1620           & PG2300  & 3750 & 1.25 & 2.5 & 5085--6040 \\
 6 February 2026  & $700\times 6$  & PG2300  & 4200 & 1.5  & 1.8 & 6033--6850 \\
 7 February 2026  & $700\times 3$  & PG2300  & 4200 & 1.5  & 1.9 & 6033--6850 \\
\hline
\end{tabular}
\end{table*}

\subsection{LAMOST spectra}
\label{sec:lamost}

The ``Large Sky Area Multi-Object Fiber Spectroscopic Telescope''
\citep[LAMOST; e.g.][]{Zhao2012} obtained two low resolution spectra of
ATL~0602.\footnote{\url{https://www.lamost.org/dr11/v1.1/search}}
The exposure time was 75~min for both of these, with approximate resolution
$R \sim 1500$. It is noteworthy that the first spectrum included an eclipse of
the WD (phase coverage 0.744--0.033) while both stars were visible throughout
for the second (phases 0.086--0.375).

The flux-calibrated spectra of the two stars are dominated by features typical
of an early to mid M dwarf, but with an excess of radiation at the shorter
wavelengths. Obviously, if a hot component of the binary is only prominent in
the blue, then it has to be smaller than the RD -- i.e. a WD.

\section{The Light Curves}
\label{sec:LCs}

\subsection{ZTF~0329}

In Fig.~\ref{fig:ztf0329lc} there
are considerable differences between the shapes of the light curves obtained in
2025, and those from earlier observation runs. Also interesting is the fact that
the amplitude is largest in $R$, followed by $I$, and then $V$. This points to
non-thermal radiation in the binary system.

Time series photometry of ZTF~0329 is available from the Catalina Sky Survey
\citep[CSS;][]{Drake2014},\footnote{\url{http://nesssi.cacr.caltech.edu/DataRelease/}}
the Asteroid Terrestrial-impact Last Alert System
\citep[ATLAS;][]{Heinze2018},\footnote{\url{http://mastwed.stsci.edu/ps1casjobs/}}
the Zwicky Transit Factory
\citep[ZTF;][]{Bellm2019},\footnote{\url{https://irsa.ipac.caltech.edu/cgi-bin/Gator/nph-scan?mission=irsa\&submit=Select\&projshort=ZTF}}
and the Near-Earth Object Wide-field Infrared Survey
\citep[NEOWISE;][]{Mainzer2014}\footnote{\url{https://irsa.ipac.caltech.edu/cgi-bin/Gator/nph-scan?mission=irsa\&submit=Select\&projshort=WISE}}
(although the signal is not unambiguously detected in the WISE $W2$ band).
From the SAAO photometry, primary eclipse times are at
\begin{equation}
T_n=2459831.60696(4.4\text{E}{-4})+0.1437540(2.3\text{E}{-7})n \;\; {\rm (d)}
\label{eq:eph_ztf}
\end{equation}
the period being the mean of the CSS, ZTF and $W1$-derived values.

\begin{figure*}
\centering
\includegraphics[width=\textwidth]{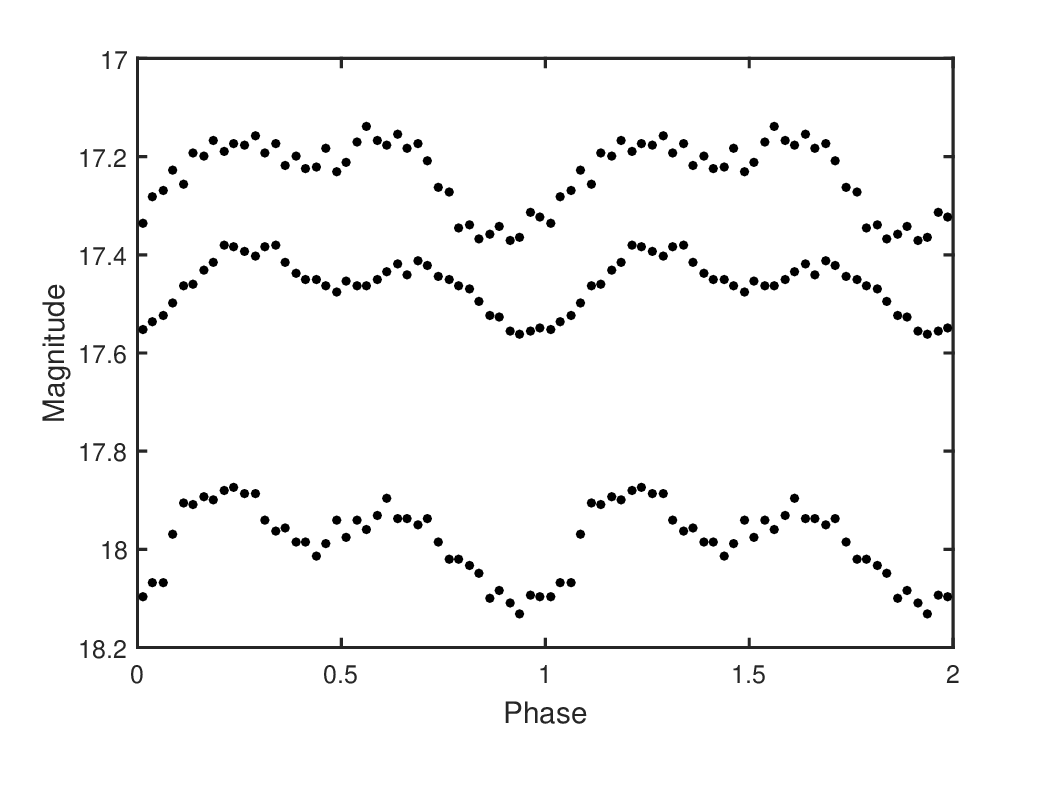}
\caption{Binned phase folded time series photometry of ZTF~0329, from
         literature sources. From top to bottom: CSS, ATLAS $o$ and ZTF $r$.}
\label{fig:ztf0329other}
\end{figure*}

Supplementary to the light curves in Fig.~\ref{fig:ztf0329lc},
Fig.~\ref{fig:ztf0329other} contains phase folded light curves from CSS, ATLAS
and ZTF, each obtained over the course of several years (see
Table~\ref{tab:ztfarch}). It is noteworthy that the ATLAS $o$ and ZTF $r$
light curves, with roughly similar bandpasses to Cousins $R$, show the same
asymmetry as the left hand panel $R$ light curve in Fig.~\ref{fig:ztf0329lc}.
This may mean that this form of variability in this spectral range is more
common than the more symmetrical shape in the right hand panel of
Fig.~\ref{fig:ztf0329lc}. It should be noted that the SALT spectra of the star
were acquired approximately contemporaneously with the light curves in the right
hand panel of Fig.~\ref{fig:ztf0329lc}, and hence may not be typical.

\begin{table}
\centering
\caption{Details of the photometry of ZTF~0329 used to produce
         Fig.~\ref{fig:ztf0329other}.}
\label{tab:ztfarch}
\begin{tabular}{cccc}
\hline\hline
Filter    & Interval covered & $N$  & Effective wavelength \\
          & (MJD)            &      & (\AA) \\
\hline
CSS       & 53644--57424     &  447 & 5628 \\
ATLAS $o$ & 57278--61030     & 2657 & 6630 \\
ZTF $r$   & 58320--60609     &  443 & 6366 \\
\hline
\end{tabular}
\end{table}

The CSS light curve of the star most closely resembles the more recent $I$ band
light curve in Fig.~\ref{fig:ztf0329lc}. This may be a result of the very
broad wavelength coverage of the Catalina
filter ($\sim 3000$--$10000$~\AA);\footnote{\url{http://svo2.cab.inta-csic.es/theory/main/}}
given the preponderance of long wavelength emission from the star (e.g.\
Fig.~\ref{fig:saltspect1}), the effective response of the filter in this case
resembles $I$ more than $R$ or $V$.

Perhaps the most striking feature of the CSS light curve is that its phase
seems slightly offset from that of the other two light curves in the Figure.
Whereas the ATLAS and ZTF observations covered roughly the same time period,
the CSS photometry largely predates them -- still, it is difficult to think of
an explanation for any phase drift in the light curve.

\begin{figure*}
\centering
\includegraphics[width=\textwidth]{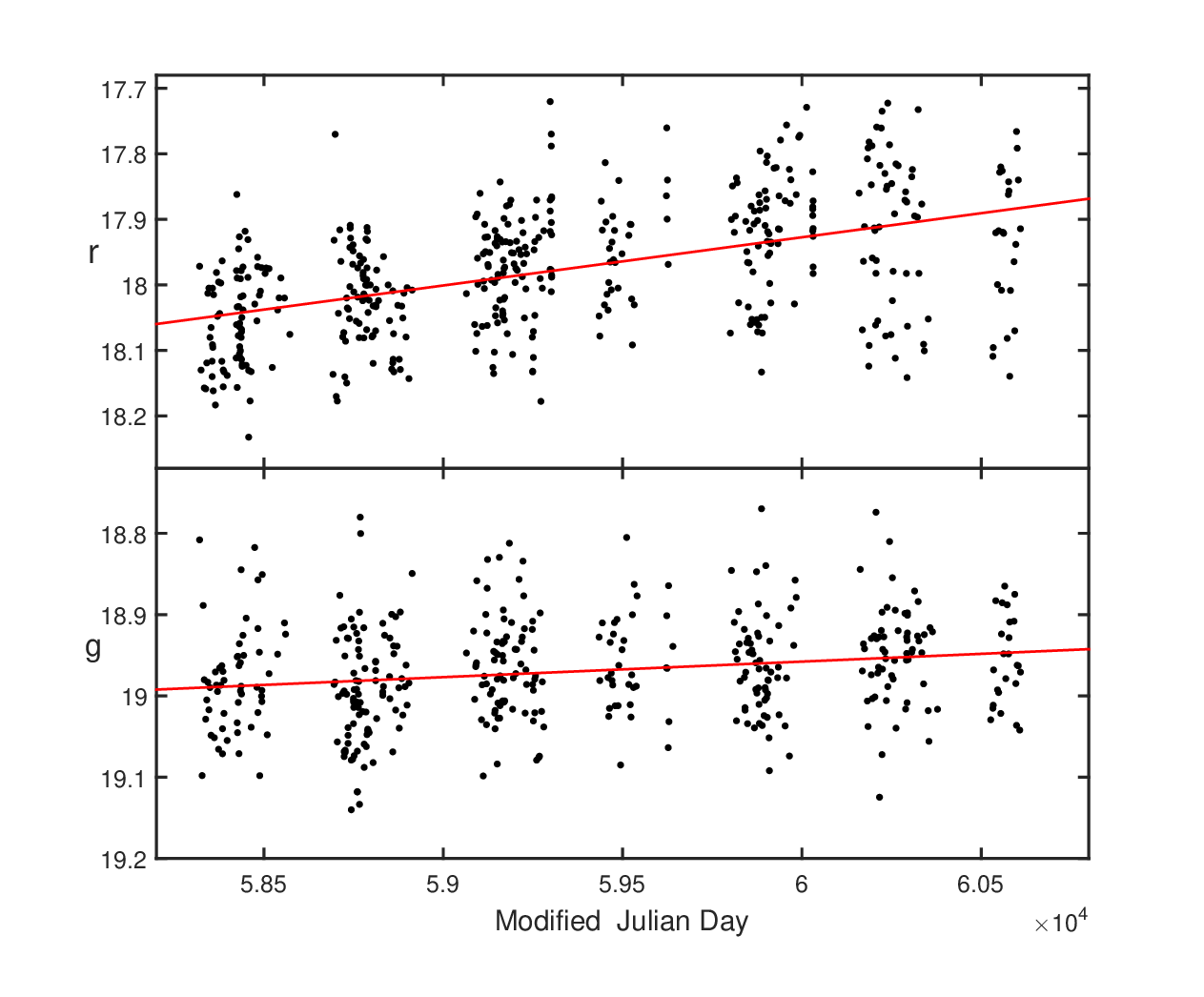}
\caption{Zwicky Transit Facility photometry of ZTF~0329. The lines are linear
         least squares fits to the magnitudes.}
\label{fig:ztf0329gr}
\end{figure*}

Linear fits to the ZTF photometry of the star (Fig.~\ref{fig:ztf0329gr}) show
highly significant brightening of $\sim 7$~mmag/yr in $g$ and $\sim 27$~mmag/yr
in $r$. (Closer scrutiny suggests a more rapid flux increase over the first four
seasons, followed by a slow decline.) Brightenings of $\sim 14$~mmag/yr and
$\sim 12$~mmag/yr are also present in the ATLAS $c$ and $o$ bands over the
period $57278<\text{MJD}<61031$.

It is therefore clear that there are both long term changes in the mean light
level of the binary, and changes in the shapes of the light curves. The
brightness change in $r$ is particularly pronounced.

\subsection{ATL~0602}

\begin{figure}
\centering
\includegraphics[width=\columnwidth]{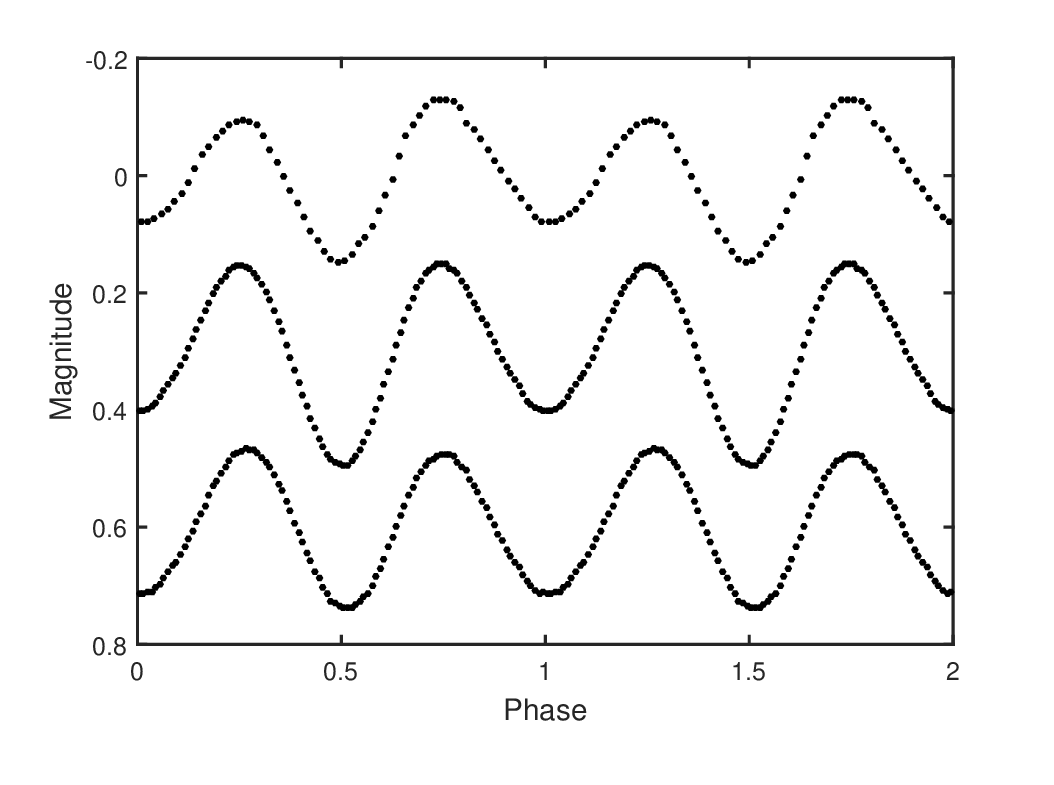}
\caption{Binned phase folded {\it TESS} photometry of ATL~0602. From top to
         bottom, sectors 6, 33 and 87. Magnitude zeropoints are arbitrary.}
\label{fig:tessfig}
\end{figure}

The phase-folded SAAO photometry of ATL~0602 is plotted in
Fig.~\ref{fig:atl0602lc}. The secondary eclipse is of similar depths in $BVR$,
but noticeably shallower in $I$. Maxima preceding primary eclipse are slightly
higher than the following ones. The light curves derived from observations by
ATLAS, CSS, WISE and ZTF are not particularly informative, but the {\it TESS}
light curves are much better sampled (Table~\ref{tab:tess}), and reveal changes
with time (Fig.~\ref{fig:tessfig}). There is a clear O'Connell effect
(different heights of the two maxima -- \citealt{OConnell1951,LiuYang2003}) in
the sector~6 light curve only. Peak-to-peak amplitudes are variable, being
0.28~mag in sector~6, 0.34~mag during sector~27 and 0.28~mag in sector~87.
Relative eclipse depths also change.

\citet{Green2023} used {\it TESS} photometry to derive the ephemeris
$$T_n=2458479.49249(0.0017)+0.196148(0.0000015)n \;\; {\rm (d)}$$
for the mid-eclipse times of the RD component in ATL~0602. A more recent
ephemeris for WD mid-eclipse times
\begin{equation}
T_n=2459206.55150(2.9\text{E}{-4})+0.1964227(1.0\text{E}{-7})n \;\; {\rm (d)}
\label{eq:eph_atl}
\end{equation}
was calculated from the SAAO photometry. Extended time series photometry of the
star was acquired by CSS, ATLAS, ZTF, WISE. Standard errors of the periods
calculated from the latter two sources are smallest -- averaging these gives the
value quoted in equation~(\ref{eq:eph_atl}).

\section{Modelling Spectral Energy Distributions (SEDs)}
\label{sec:sed}

The standard approach to fitting SEDs to photometry is to convert
magnitudes to fluxes and then to compare these to model fluxes. 
It appears to the authors more natural to work directly with the
magnitudes. This also makes trivial the application of 
reddening corrections from published maps. The direct determination 
of the stellar luminosities is incorporated by relating all observed
magnitudes to the bolometric magnitudes of the two stars, through bolometric corrections. 
The unknowns to be soved are then the bolometric
magnitudes, temperatures and gravities of the stars (the latter two
parameters through the best-fitting bolometric corrections). See e.g.
\citet{KoenKniazev2024} for an application of the method.

Sources of photometry used here are {\it Gaia} \citep{GaiaCollab2023}, the
Panoramic Survey Telescope and Rapid Response System
\citep[Pan-STARRS;][]{Chambers2016}, the SkyMapper Southern Survey
\citep{Wolf2018}, the Two Micron All-Sky Survey
\citep[2MASS;][]{Skrutskie2006}, the Wide-field Infrared Survey Explorer
\citep[WISE;][]{Wright2010}, the Galaxy Evolution Explorer
\citep[GALEX;][]{Bianchi2017} and the Sloan Digital Sky Surveys
\citep[SDSS;][]{Ahumada2020}. Photometry of the stars from the various surveys
is conveniently obtainable from the VizieR
service\footnote{\url{https://vizier.cds.unistra.fr/viz-bin/VizieR}} of the
Strasbourg astronomical Data Center.

The apparent magnitude measured through a filter with effective wavelength
$\lambda$ is
\begin{eqnarray}
m_\lambda &=& -2.5\log \left [(L_1+L_2)_\lambda /L_\odot \right ]
+5(\log D -1) +A_\lambda \nonumber\\
&=&-2.5\log \left [10^{-0.4 M_{1\lambda}}
+10^{-0.4M_{2\lambda}} \right ]\nonumber\\
&&+5(\log D -1)+A_V f_\lambda \nonumber\\
&=&-2.5\log\left [10^{-0.4 (M_{1\mathrm{bol}}-BC_{1\lambda})}
+10^{-0.4(M_{2\mathrm{bol}}-BC_{2\lambda})} \right ]\nonumber\\
&&+5(\log D -1)+A_V f_\lambda\nonumber\\
& \equiv & m_{\lambda 0}+A_V f_\lambda
\label{eq:mag}
\end{eqnarray}
where subscripts 1 and 2 refer to the two stars respectively, $D$ is the
distance (in pc), $M_{j\lambda}$ is the monochromatic absolute magnitude of
star $j$, and $M_{j\mathrm{bol}}$ and $BC_{j\lambda}$ are its bolometric
magnitude and bolometric correction.
Distances $d$ are conveniently available from e.g.\ \citet{BailerJones2021}.
The last term in equation~(\ref{eq:mag}) describes the effect of interstellar
absorption $A_\lambda=A_V f_\lambda$. The function $f_\lambda$ is available
from table~3 in \citet{WangChen2019}, supplemented by \citet{Casagrande2019}
for SkyMapper and \citet{Wall2019} for {\it GALEX} filters. Interstellar
absorption $A_V$ can be determined from the high resolution G-Tomo reddening
maps\footnote{\url{https://explore-platform.eu/sdas/about/gtomo}}
\citep{Vergely2022,Lallement2022} or estimated from the photometry (see below).

Bolometric corrections are available from e.g.\ the ``MESA Isochrones and
Stellar Tracks'' website\footnote{\url{http://waps.cfa.harvard.edu/MIST/model_grids.html\#bolometric}}
\citep{Choi2016}. Stellar atmosphere models and theoretical spectra are used
in the calculation of the corrections, which also incorporate filter
transmission curves \citep[see][for the pertinent equations]{Girardi2008}.
The \citet{Choi2016} bolometric corrections are supplied in tables for ranges
of values of effective temperature, gravity, metallicity and dust extinction.
In the present implementation metallicity is assumed to be solar, and reddening
is modelled explicitly.

Equation~(\ref{eq:mag}) depends explicitly on the bolometric magnitudes, and,
through its dependence on bolometric corrections, implicitly on the temperatures
and gravities of the two stars. These six parameters can be determined by
minimising
\begin{equation}
SS=\sum_\lambda [m_\lambda({\rm theoretical})-
m_\lambda({\rm observed})]^2\equiv \sum_\lambda r^2_\lambda
\label{eq:ss}
\end{equation}
where $r_\lambda$ are the model residuals. Standard errors on estimated
quantities can be calculated by bootstrapping, using the model residuals.
These errors are bound to be too optimistic, as they do not take into account
uncertainties in bolometric corrections and interstellar extinctions, and ignore
the fact that there may be many locally optimal sets of parameter solutions.

Results for the two stars are given in Table~\ref{tab:sed}. The SEDs are clearly dominated by the RDs, hence all the WD
parameters are highly uncertain (particularly so for ATL~0602).
The WD temperatures are much better determined from 
spectral continuum slopes -- see the next section of the paper.

\begin{figure}
\centering
\includegraphics[width=\columnwidth]{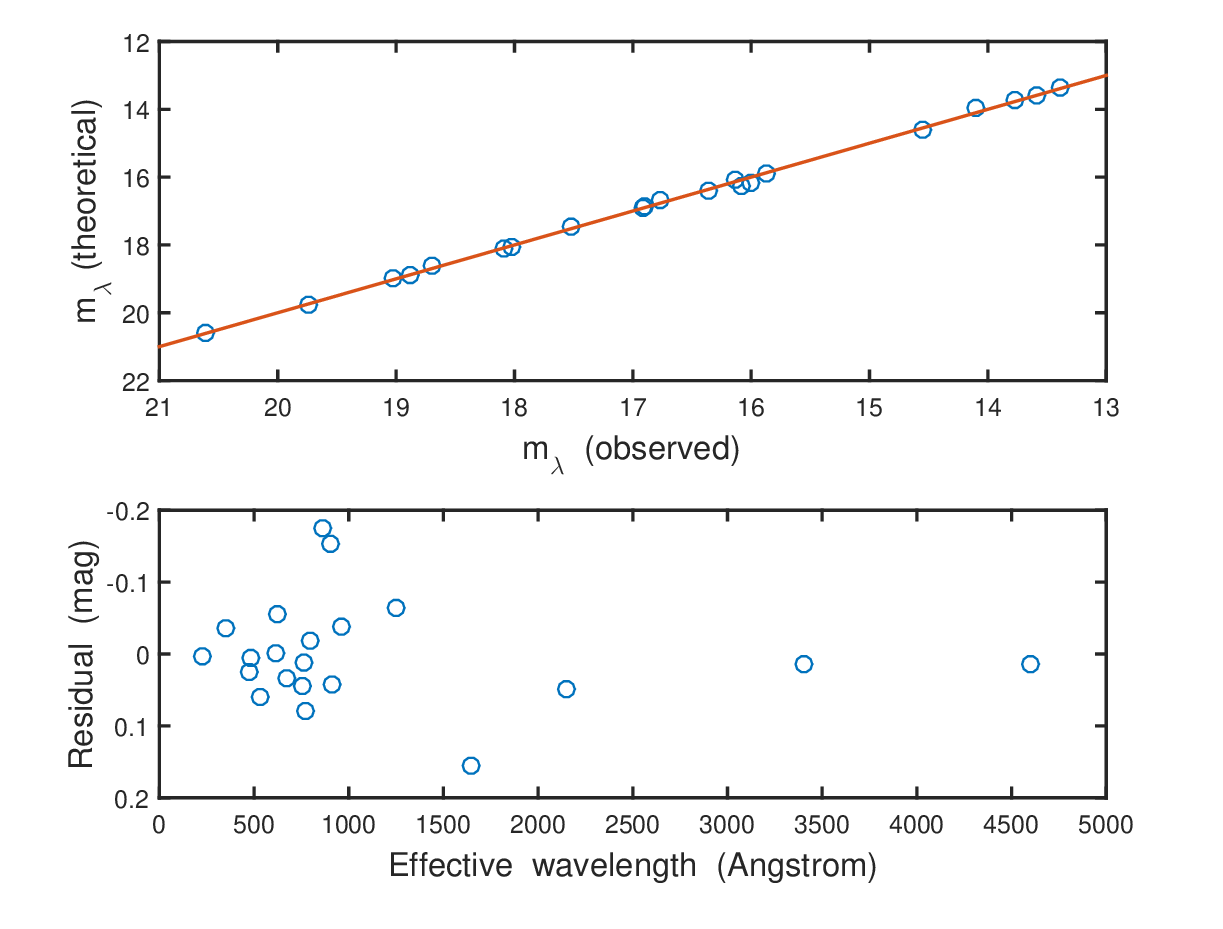}
\caption{Top panel: a comparison between the observed and theoretical [from Equation (3)]
         apparent magnitudes of ZTF~0329.
         Bottom panel: The residuals, as defined in equation~(\ref{eq:ss}), from fitting a
         SED to available standardised photometry of ZTF~0329.}
\label{fig:sed0329}
\end{figure}

\begin{figure}
\centering
\includegraphics[width=\columnwidth]{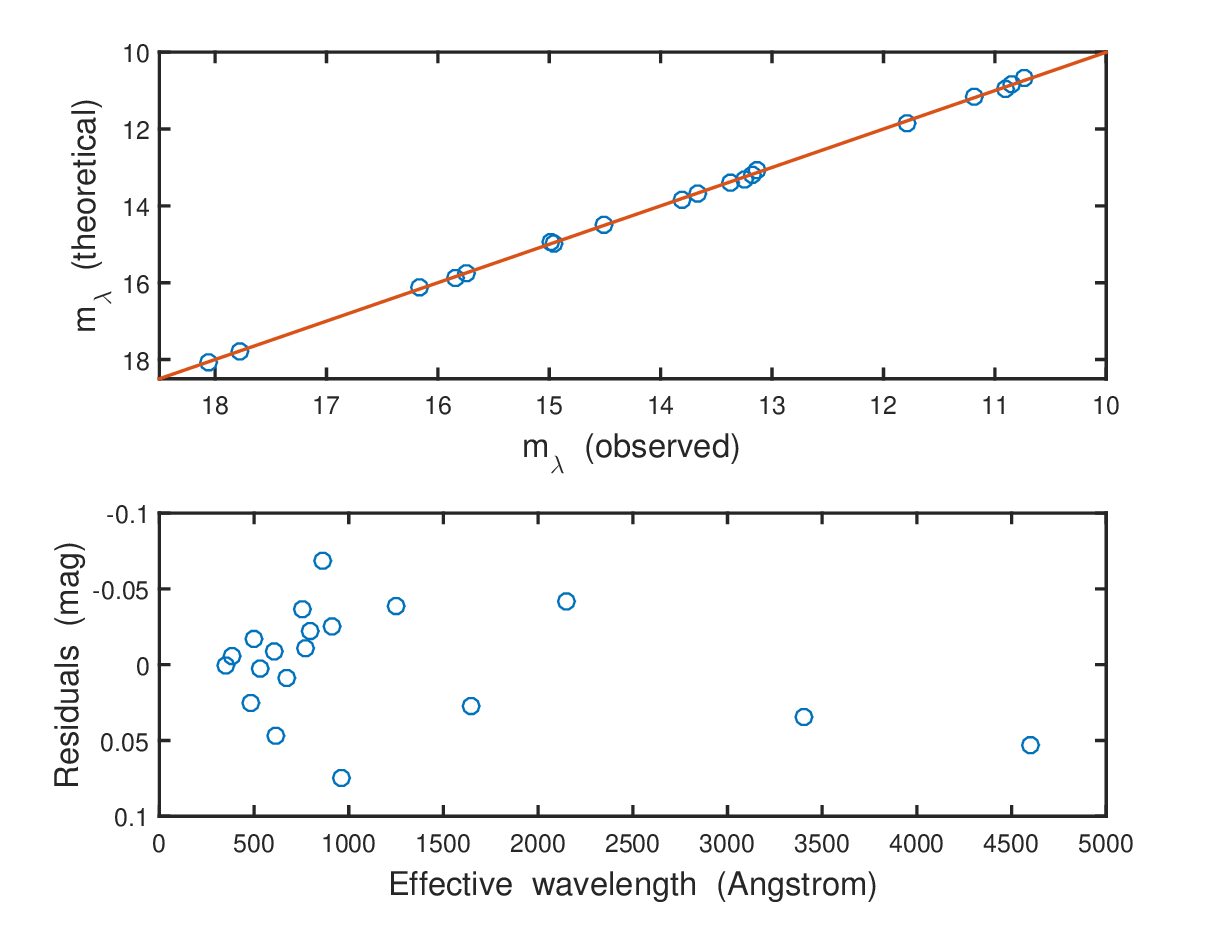}
\caption{As for Fig.~\ref{fig:sed0329}, but for ATL~0602.}\
\label{fig:sed0602}
\end{figure}

Figs.~\ref{fig:sed0329} and \ref{fig:sed0602} show the model fits and 
residuals $r_\lambda$ for the two stars. Note that the vertical scales of
the two residual plots differ by a factor two. This is due primarily to
two bright outliers in the ZTF~0329 photometry
at PanSTARRS $z$ ($\lambda_{\rm eff} \sim 8660$~\AA) and SDSS
$z^\prime$ ($\lambda_{\rm eff} \sim 9050$~\AA). It is tempting to speculate
that this could be due to excess emission in this wavelength range -- but then
SkyMapper $z$ ($\lambda_{\rm eff} \sim 9121$~\AA) is $\sim 0.2$~mag fainter.
There is also a faint outlier at 2MASS $H$ ($\lambda \sim 16500$~\AA).
Table~\ref{tab:sed} also contains a second set of model parameters for ZTF~0329
calculated without the three outlying datapoints.

Equations~(\ref{eq:mag}) and (\ref{eq:ss}) can also be used to estimate the
extinction most compatible with the photometry:
\begin{equation}
A_V=\max \left \{ \begin{array}{ll}
 \displaystyle\sum_\lambda [m_\lambda({\rm observed})-
m_{\lambda 0} ]f_\lambda \left / \sum_\lambda f^2_\lambda \right. \\
   \\
0
\end{array} \right .
\label{eq:av}
\end{equation}
For the full ZTF~0329 dataset, and for ATL~0602, the optimal value is
$A_V=0$~mag. Not surprisingly, the RMS of the residuals is smaller with $A_V$
estimated from equation~(\ref{eq:av}), rather than taken from the reddening
maps. A glance at Table~\ref{tab:sed} shows that the estimated parameters based
on the different values of $A_V$ are nonetheless in excellent agreement.

The error budget associated with Equation (2) is interesting. Uncertainties in the 
theoretical magnitudes include those due to 
errors in $A_V$, uncertainties in the bolometric corrections
(both due to modelling uncertainties and the estimation errors in 
the temperatures and gravities), and the distance error -- see (1).
Easiest to quantify is the effect of the uncertainty in the distance,
which, by the delta method, is approximately $2.17 \sigma_D/D$; this
is negligible for ATL~0602 (0.006 mag), but cannot ignored for the more distant ZTF~0329 (0.043 mag). 

As far as the observed magnitudes are concerned, not all
photometric catalogues supply estimates of the uncertainties 
on their published magnitudes. Contributions due to the intrinsic
variability of the stars will depend on the
phases at which measurements are taken, so it is not clear how reliable
observational error estimates for these stars are. Available
errors lie in the range 0.013-0.051 mag for ATL~0602 (median 0.023 mag);
for ZTF~0329 errors for optical/IR are 0.004-0.081 mag, while the
error on the GALEX NUV measurement is 0.199 mag (median 0.042 mag).

The RMS values  
$$\sigma^2=\left [\frac{1}{N}\sigma_\lambda r_\lambda^2 \right ]$$ 
in Table~\ref{tab:sed} can be roughly placed in context by referring to
the discussion in the preceding two paragraphs. Let
$$\sigma_*=\sqrt{\sigma_1^2+\sigma_2^2} \; ,$$
where $\sigma_1$ is uncertainty in 
$m_\lambda$(theoretical) due to $\sigma_d$, and $\sigma_2$ is the median 
error in $m_\lambda$(observed). The ratio
$$r=\frac{N \sigma^2}{(N-p) \sigma^2_*}$$
is of the form of a reduced $\chi^2$ statistic, adjusted for the 
number $p$ of estimated parameters --see the last column of Table~\ref{tab:sed}.

Values of $r$ should be seen as order of magnitude reassurance that
results are reasonable, rather than taken at their numerical face
values. For example, the statitics for ATL~0602 are changed from
3.30 and 3.341 to 1.14 and 1.18 if $\sigma_2=0.04$ mag -- quite
reasonable bearing in mind the large variability (amplitudes 
$\gtrsim 0.29$ mag) in the star -- see Fig. 2.

The small errors on the temperatures and luminosities of the RD components
allow accurate calculation of their radii:
\begin{eqnarray}
R &=&(T_\odot/T)^2 10^{0.2(M_{\mathrm{bol}\odot}-M_{\mathrm{bol\,RD}})}\nonumber\\
&=& (5774/T)^2 10^{0.2(4.74-M_{\mathrm{bol\,RD}})}
\label{eq:radius}
\end{eqnarray}
The radii in Table~\ref{tab:sed} can be compared to those of typical M dwarfs
with similar effective
temperatures\footnote{\url{http://www.pas.rochester.edu/~emamajek/EEM_dwarf_UBVIJHK_colors_Teff.txt}}
\citep{Pecaut2012,PecautMamajek2013}. For ZTF~0329 and ATL~0602 the expected
radii are $R \sim 0.26\,R_\odot$ and $R \sim 0.36\,R_\odot$ respectively: the
radius of the RD in ZTF~0329 is completely typical whereas the M dwarf radius
in ATL~0602 is inflated by almost 60 per cent.

\begin{table*}
\centering
\caption{Optimal SED models fitted to standardised photometry of the two stars.
         Standard errors of estimates are given in brackets. The 
penultimate column
         contains the residual standard deviation and the last column
a crude goodness-of-fit statistic. Two sets of solutions are
         presented for ZTF~0329, respectively for the full dataset ($N=21$),
         and for a trimmed dataset ($N=18$) with three outlying measurements
         excluded. For each dataset, models with reddening respectively from
         extinction maps (first line), and estimated from the photometry
         (second line), are presented.}
\label{tab:sed}
\begin{tabular}{ccccccccccccc}
\hline\hline
  && \multicolumn{4}{c}{Red dwarf properties} &&
\multicolumn{3}{c}{White dwarf properties}  & &\\
$N$ & $A_V$ & $T_{\rm eff}$ & $\log\,g$ & $M_{\rm bol}$ & $R$ &&
$T_{\rm eff}$ & $\log\,g$ & $M_{\rm bol}$ & $\sigma$ & $r$\\
 & (mag) &  (K) & & (mag) & ($R_\odot$) && (K) & & (mag) & (mag) & \\
\hline
\multicolumn{11}{c}{ZTF~0329}\\
21 & 0.19       & 3200(40)   & 4.7(0.47) & 10.22(0.04) & 0.261(0.008) && 9987(415)   & 9.5(0.62) & 12.71(0.08) & 0.0711 & 1.96\\
   & 0.00(0.09) & 3156(37)   & 5.0(0.47) & 10.28(0.05) & 0.262(0.007) && 9546(451)   & 9.5(0.55) & 13.00(0.17) & 0.0681 & 1.93\\
\hline
18 & 0.19       & 3179(17)   & 5.2(0.28) & 10.23(0.04) & 0.263(0.005) && 9828(194)   & 9.5(0.36) & 12.67(0.05) & 0.0369 & 0.57\\
   & 0.16(0.09) & 3175(29)   & 5.1(0.31) & 10.24(0.05) & 0.262(0.005) && 9765(324)   & 9.5(0.39) & 12.71(0.14) & 0.0368 & 0.61\\
\hline
\multicolumn{11}{c}{ATL~0602}\\
19 & 0.03       & 3373(12)   & 5.5(0.18) & 8.29(0.01)  & 0.572(0.006) && 9020(12450) & 9.5(0.94) & 12.62(0.77) & 0.0357 & 3.30\\
   & 0.00(0.06) & 3364(22)   & 5.5(0.22) & 8.30(0.02)  & 0.573(0.006) && 9063(13410) & 9.5(1.07) & 12.68(0.85) & 0.0349 & 3.41\\
\hline
\end{tabular}
\end{table*}

\section{Modelling of the Low Resolution Spectra}
\label{sec:specmod}

We fit models to the two SALT low resolution spectra of ZTF~0329 with the
highest signal-to-noise ratios, as well as the two LAMOST spectra of ATL~0602.
Each spectrum $S$ was modelled as the sum of two spectra, $S_w$ and $S_r$
representing a WD and RD respectively:
\begin{equation}
S(\lambda)=\alpha S_w (\lambda)+\beta S_r(\lambda)
\label{eq:specmod}
\end{equation}
where $0 \le \alpha ,\beta \le 1$ are scaling factors.

The best-fitting combination was determined by performing a grid search over
libraries of spectra. Gaussian convolution was used to reduce the resolution of
the library spectra to that of the SALT or LAMOST spectra. For the WD spectra,
a single library source was
used\footnote{\url{http://svo2.cab.inta-csic.es/theory/newov2/index.php?models=koester2}}
-- see \citet{Tremblay2009} and \citet{Koester2010} for details. Two sources
were used for RD spectral templates -- theoretical NextGen solar composition
models\footnote{\url{http://svo2.cab.inta-csic.es/theory/newov2/index.php?models=NextGen}}
\citep{Allard1997,Hauschildt1999} and empirical templates constructed from SDSS
spectra\footnote{\url{http://svo2.cab.inta-csic.es/theory/newov2/index.php?models=tpl_kesseli}}
\citep{Kesseli2017}. Chemical composition was assumed solar.

\begin{figure}
\centering
\includegraphics[width=\columnwidth]{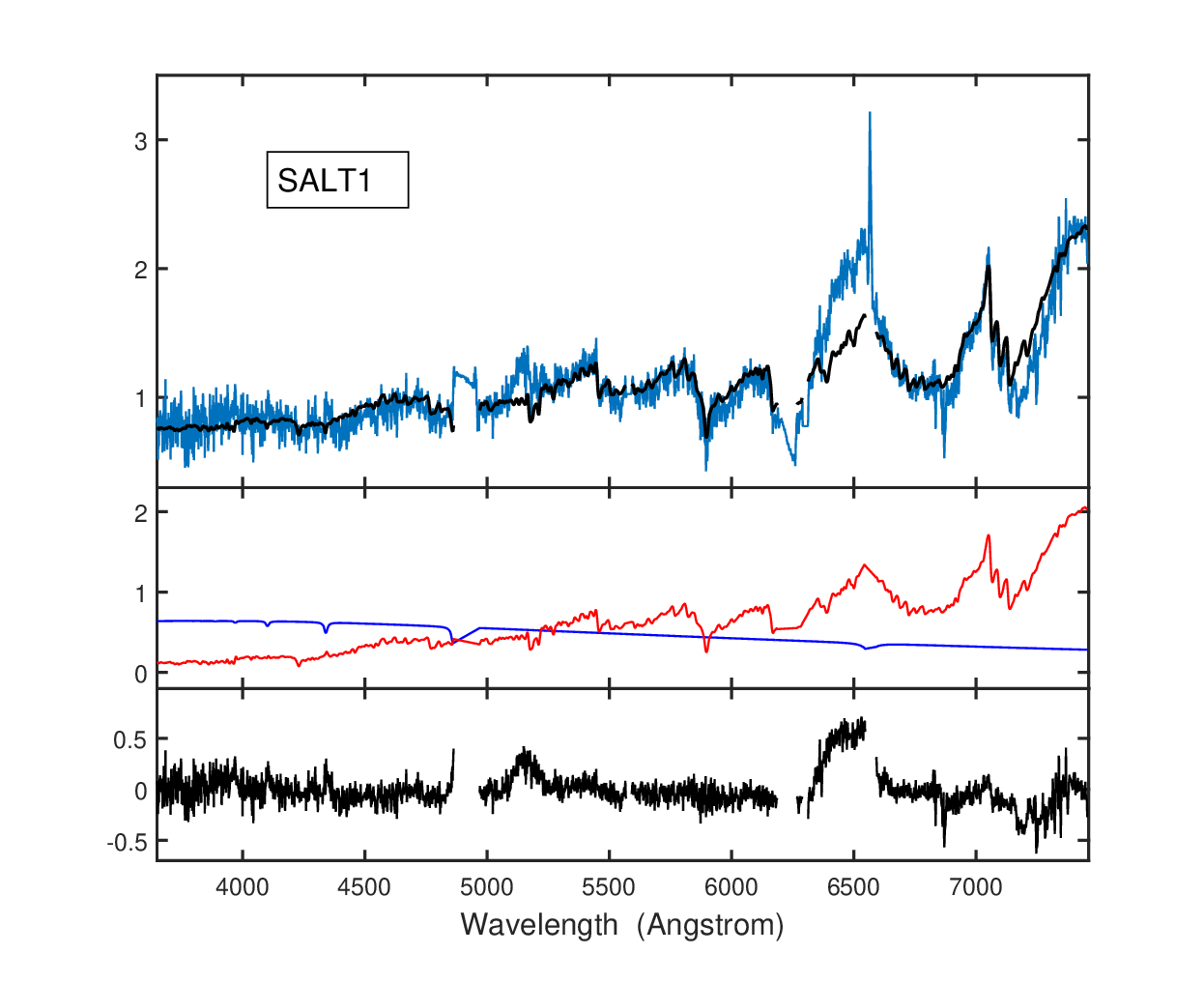}
\caption{Top panel: First of the two best SALT low resolution spectra of ZTF~0329 (blue) 
with the optimal model fit (black). The binary phase
covered by the observation was 0.867-0.939.
Middle panel: The RD (red line) and WD (blue line) component
spectra of the model.
Bottom panel: The residuals from the fitted model.}
\label{fig:saltspect1}
\end{figure}

\begin{figure}
\centering
\includegraphics[width=\columnwidth]{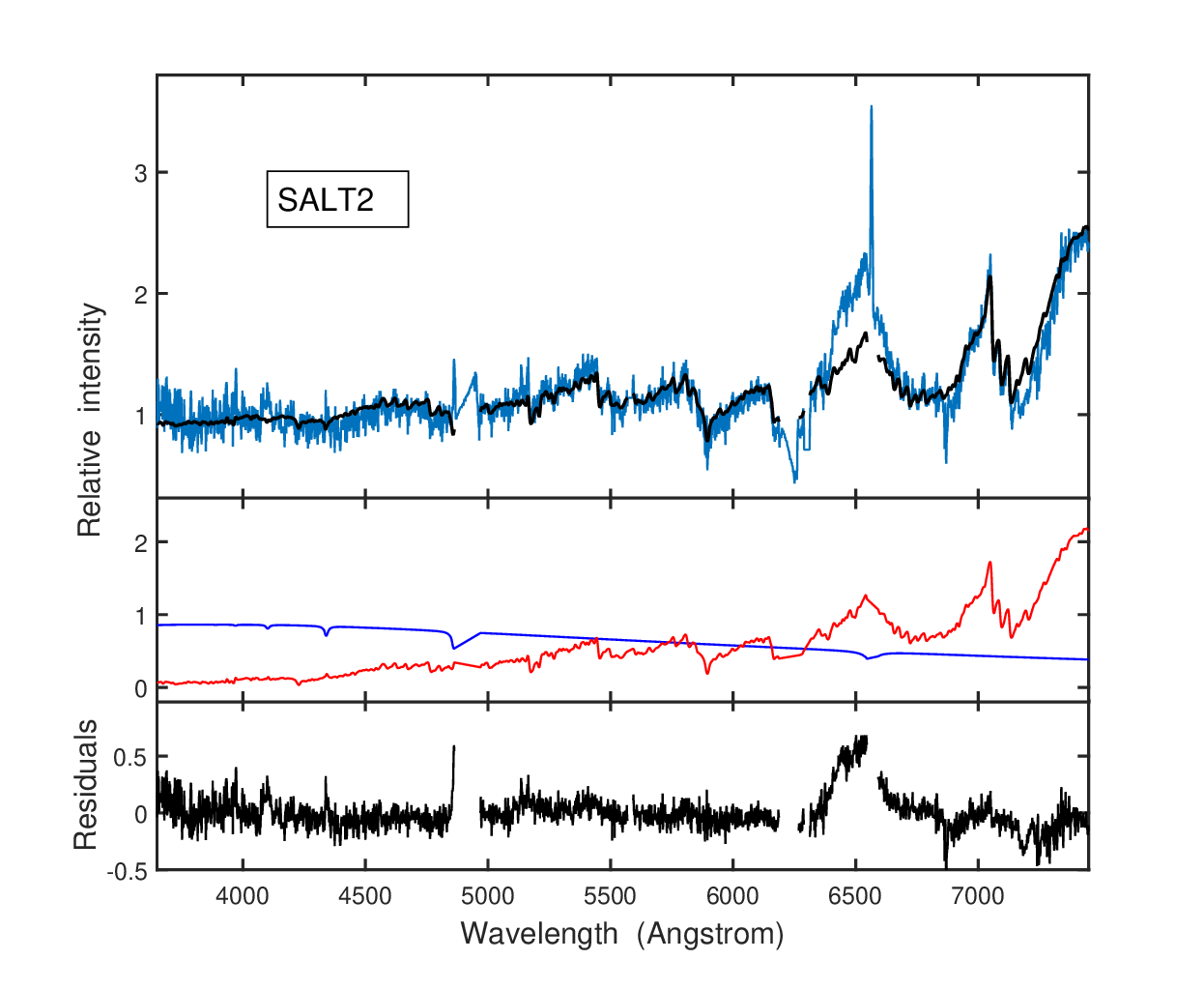}
\caption{As for Fig.~\ref{fig:saltspect1}, but for the second of the best SALT spectra of ZTF~0329, covering 
the binary phase interval 0.941-0.013.}
\label{fig:saltspect2}
\end{figure}

\begin{figure}
\centering
\includegraphics[width=\columnwidth]{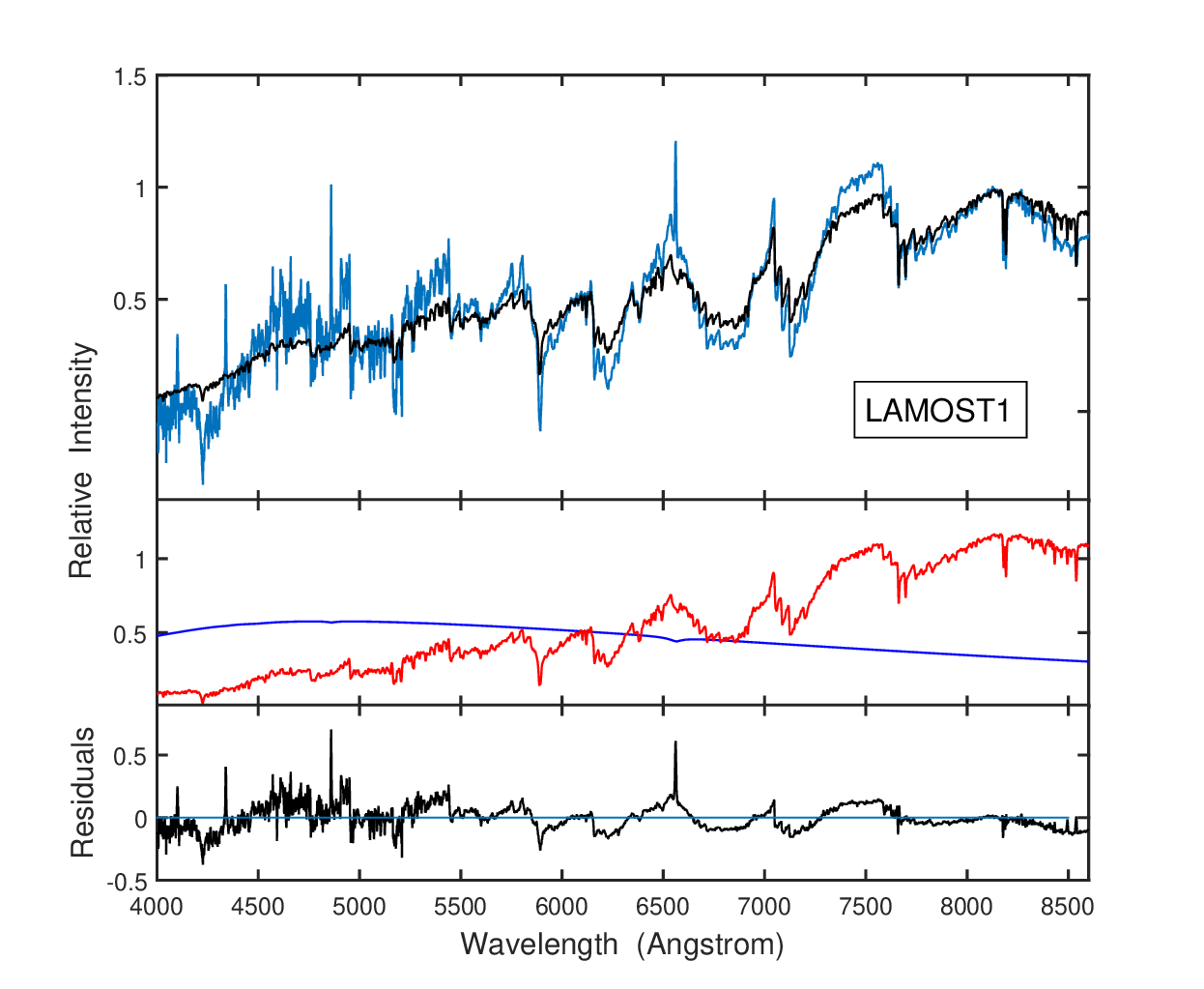}
\caption{As for Fig.~\ref{fig:saltspect1}, but showing the first LAMOST spectrum of ATL~0602,
covering the binary phase interval 0.744-0.033.}
\label{fig:lamost1}
\end{figure}

\begin{figure}
\centering
\includegraphics[width=\columnwidth]{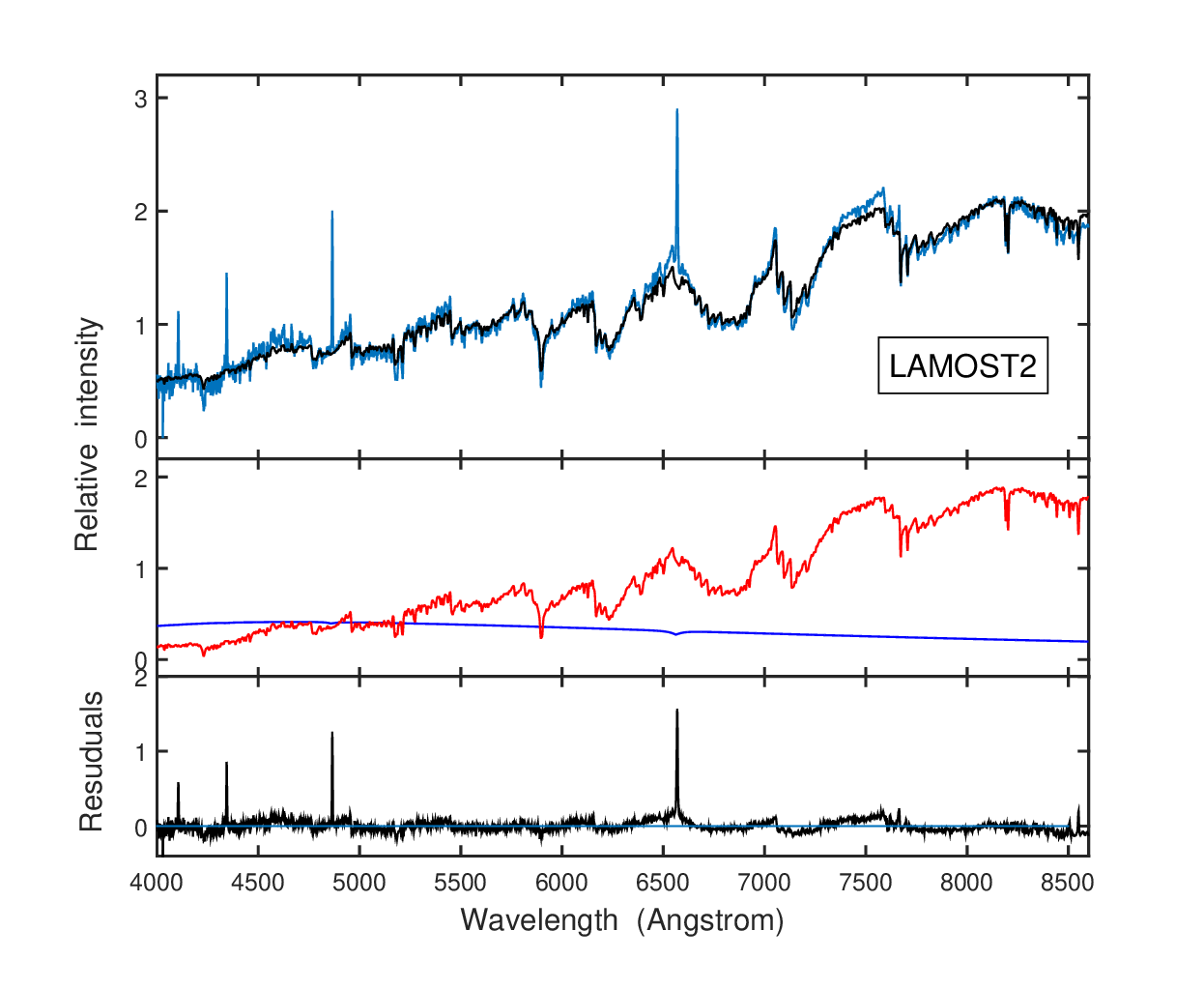}
\caption{As for Fig.~\ref{fig:saltspect1}, but showing the first LAMOST spectrum of ATL~0602,
covering the binary phase interval 0.086-0.375.}
\label{fig:lamost2}
\end{figure}

The spectral modelling results can be found in Table~\ref{tab:specmod}, with
accompanying plots in Figs.~\ref{fig:saltspect1}--\ref{fig:lamost2}. The
fit to the first LAMOST spectrum is by far the worst -- this may be due to the
fact that the exposure covered an eclipse of the WD. A glance at
Fig.~\ref{fig:atl0602lc} shows that this would have had a particularly
pronounced effect on the blue part of the spectrum.

In order to be able to attach some meaning to the weights $\alpha$ and $\beta$
in equation~(\ref{eq:specmod}), all spectra were normalised by their fluxes at
$\lambda=5550$~\AA. The last column of Table~\ref{tab:specmod} contains
$$RMS=\left \{ \frac{1}{N_\lambda} \sum_\lambda 
[S(\lambda)-\alpha S_w(\lambda)-\beta S_r(\lambda)]^2
 \right \}^{1/2}$$
where $N_\lambda$ is the number of wavelength elements in the spectrum
It is clear that by this token the empirical template RD spectra fit the
observed spectrum better than the theoretical spectra.
According to the \citet{Pecaut2012} and \citet{PecautMamajek2013}\footnote{
\url{http://www.pas.rochester.edu/~emamajek/EEM_dwarf_UBVIJHK_colors_Teff.txt}}
tables, $T_{\rm eff}=3410$~K for a typical M3 dwarf, in excellent agreement
with the 3400~K temperature from the NextGen model fitted to spectra of
ATL~0602. For ZTF~0329, the temperatures corresponding to M4--M5 spectral
types are 3200--3030~K,\footnote{\url{http://www.pas.rochester.edu/~emamajek/EEM_dwarf_UBVIJHK_colors_Teff.txt}}
which can be compared to the NextGen temperatures of 3000 and 3100~K.

The temperatures of the WD components derived from different spectra agree
quite well, but depend on which model is used for the RD. Given the superior
fit of the empirical template spectra, those models are preferred, i.e. WD
temperatures of 7000~K (ZTF~0329) and 5750--6000~K (ATL~0602).

The model fitting favours WD empirical models with very high gravities
in which the Hydrogen absorption lines are largely washed
out by the intense broadening, which agrees with the observed spectra.  
High signal-to-noise observations at higher resolution would reveal
the extent to which the WD absorption wings are present, thus placing
realistic limits on the gravity of the WD and its Hydrogen abundance.

RD properties were also determined using the FBS package (see next section);
for ZTF~0329 we obtained $T_{\rm eff}=2965(30)$~K, $\log\,g=4.54(0.03)$, and
for ATL~0602, $T_{\rm eff}=3360(50)$~K and $\log\,g=4.82(0.15)$. ZTF~0329 is
metal-rich ([Fe/H]=$0.18\pm0.05$) while ATL~0602 is slightly metal-poor
([Fe/H]=$-0.30\pm0.05$).

\begin{table*}
\centering
\caption{A summary of models fitted to the classification spectra of the two
         stars. The SALT and LAMOST spectra were taken of ZTF~0329 and
         ATL~0602 respectively. The acronyms ``ET'' and ``NGS'' represent
         the ``empirical template'' and ``NextGen solar'' spectra $S_r$
         respectively. Spectra in the former library are given for various
         spectral types, rather than temperature and gravity. The last column
         shows the root mean square of the residuals.}
\label{tab:specmod}
\begin{tabular}{ccccccccccc}
\hline\hline
  & & & & & \multicolumn{2}{c}{Red dwarf} &&
\multicolumn{2}{c}{White dwarf} & \\
Red spectrum & $\alpha$ & $\beta$ & $V_r$ && $T$ & $\log\,g$ &&
 $T$ & $\log\,g$ & $RMS$ \\
   & & & (km\,s$^{-1}$) & & (K) & && (K) & &\\
\hline
\multicolumn{11}{c}{SALT1}\\
ET  & 0.48 & 0.56 & 156    && \multicolumn{2}{c}{M4} && 7000 & 9.25 & 0.171 \\
NGS & 0.54 & 0.70 & 198    && 3100 & 4.5 && 6250 & 8.0  & 0.199 \\
\hline
\multicolumn{11}{c}{SALT2}\\
ET  & 0.65 & 0.42 &  41    && \multicolumn{2}{c}{M5} && 7000 & 9.5  & 0.153 \\
NGS & 0.65 & 0.66 &  86    && 3000 & 4.0 && 6250 & 8.0  & 0.194 \\
\hline
\multicolumn{11}{c}{LAMOST1}\\
ET  & 0.55 & 0.38 & $-178$ && \multicolumn{2}{c}{M3} && 5750 & 9.5  & 0.096 \\
NGS & 0.58 & 0.49 & $-114$ && 3400 & 5.5 && 5250 & 9.5  & 0.102 \\
\hline
\multicolumn{11}{c}{LAMOST2}\\
ET  & 0.38 & 0.59 &  201   && \multicolumn{2}{c}{M3} && 6000 & 9.5  & 0.067 \\
NGS & 0.54 & 0.70 &  260   && 3400 & 5.5 && 5000 & 6.5  & 0.135 \\
\hline
\end{tabular}
\end{table*}

\section{Radial Velocities}
\label{sec:rv}

We utilized the FBS package \citep[Fitting Binary Stars;][]{Kniazev2020},
specifically developed for the analysis of binary star system spectra and used
for spectra of different resolutions, including LAMOST low resolution data
\citep{Kniazev2023,Kniazev2025}.
The package uses \textsc{phoenix} models \citep{Husser2013} as stellar spectrum
library, as it contains spectra of low-mass stars down to an effective
temperature of $T_{\rm eff}=2300$~K and a $\log\,g$ up to $+6.0$.

Since FBS allows any number of spectral regions to be selected for analysis,
we excluded the gaps between CCDs, as well as the regions containing emission
lines and the bright regions of the night sky, from our analysis. In all cases,
when using the FBS programme to measure the errors in the output parameters, we
employed a Monte Carlo method, performing several hundred iterations, with the
final parameter values determined from the spread of values as the mean and the
root mean square (RMS) error. To speed up the process, we use task
parallelisation with the \textsc{parallel} utility \citep{Tsange2025}, and have
additionally integrated GPU support into the programme based on the
\textsc{cupy}\footnote{\url{https://github.com/cupy/cupy/}} library, which
significantly speeds up the process.

\begin{figure}
\centering
\includegraphics[width=\columnwidth]{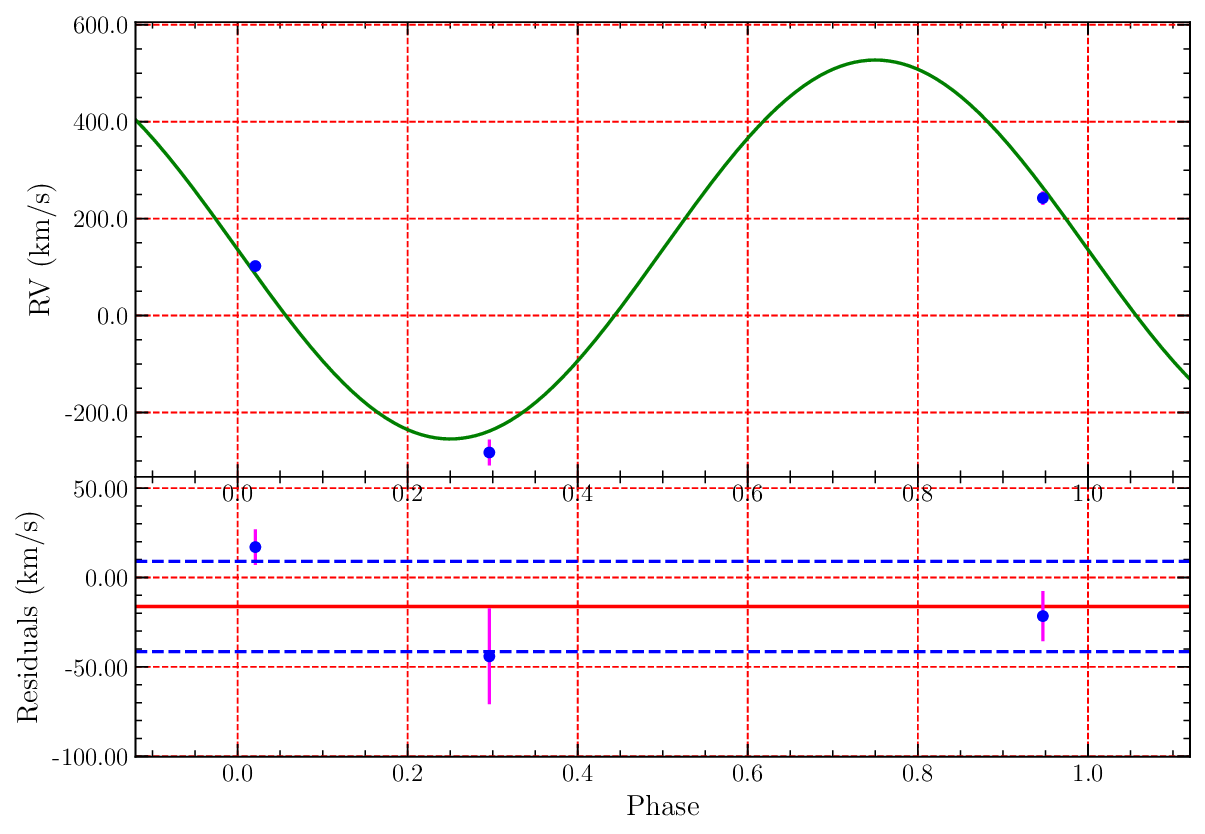}
\caption{Top panel: Phased radial velocities of ZTF~0329 (dots,
         Table~\ref{tab:rv}) with the model fitted by the FBS software (curve,
         Table~\ref{tab:rvcurve}). Bottom panel: dots indicate the residual
         values, and the blue broken lines show $\pm 1$ RMS deviations from
         zero.}
\label{fig:ztfvel}
\end{figure}

\begin{figure}
\centering
\includegraphics[width=\columnwidth]{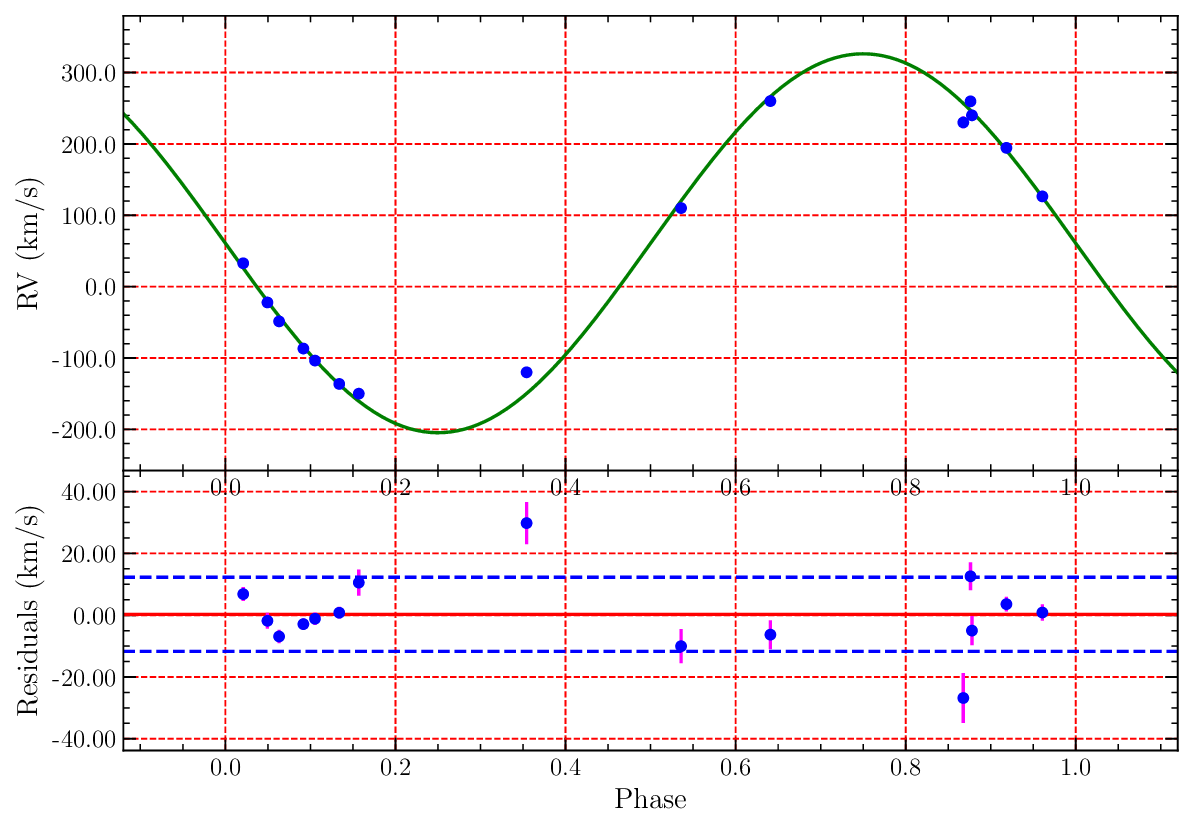}
\caption{As for Fig.~\ref{fig:ztfvel}, but for ATL~0602.}
\label{fig:atlvel}
\end{figure}

As may have been anticipated from the spectra in Figs.~\ref{fig:saltspect1}--
\ref{fig:lamost2}, only RD radial velocities can be determined.
The results are plotted in Figs~\ref{fig:ztfvel} and~\ref{fig:atlvel}, with
numerical values in Table~\ref{tab:rv}. Note that velocities determined from
the H$\alpha$ lines in the PG2300 spectra of ZTF~0329 do not phase with the
velocities determined from the full PG0700 spectra, and are therefore not
included. The model results are summarised in Table~\ref{tab:rvcurve}.

\begin{table*}
\centering
\caption{Observed and model radial velocities of the RD components of the
         binaries, determined using the FBS package.}
\label{tab:rv}
\begin{tabular}{ccrrr}
\hline\hline
BJD           & Phase & $V_{\rm hel}$  & $V_{\rm model}$ & Residual \\
(d)           &       & (km\,s$^{-1}$) & (km\,s$^{-1}$)  & (km\,s$^{-1}$) \\
\hline
\multicolumn{5}{c}{ZTF~0329}\\
2461044.35826 & 0.296 & $-282.544\pm 26.787$ & $-238.4321$ & $-44.1119$ \\
2461047.32692 & 0.947 & $  242.457\pm 13.937$ & $ 264.0500$ & $ -21.5930$ \\
2461047.33754 & 0.021 & $  102.132\pm  9.920$ & $  85.1223$ & $  17.0097$ \\
\hline
\multicolumn{5}{c}{ATL~0602}\\
2460303.38450 & 0.536 & $ 110.072\pm 5.505$ & $ 121.196$ & $ -11.124$ \\
2460331.34406 & 0.878 & $ 240.052\pm 4.658$ & $ 242.753$ & $  -2.701$ \\
2460363.31480 & 0.641 & $ 260.037\pm 4.658$ & $ 266.586$ & $  -6.549$ \\
2460396.25790 & 0.354 & $-120.060\pm 6.775$ & $-149.775$ & $  29.715$ \\
2460630.55433 & 0.157 & $-150.023\pm 4.234$ & $-161.888$ & $  11.865$ \\
2460639.53311 & 0.868 & $ 230.061\pm 8.046$ & $ 255.527$ & $ -25.466$ \\
2461078.34854 & 0.876 & $ 259.662\pm 4.475$ & $ 246.994$ & $  12.668$ \\
2461078.35684 & 0.918 & $ 194.467\pm 2.373$ & $ 190.997$ & $   3.469$ \\
2461078.36514 & 0.961 & $ 126.462\pm 2.600$ & $ 125.811$ & $   0.651$ \\
2461078.38256 & 0.049 & $  -22.194\pm 2.545$ & $ -20.203$ & $  -1.990$ \\
2461078.39086 & 0.092 & $  -86.767\pm 1.764$ & $ -83.836$ & $  -2.930$ \\
2461078.39915 & 0.134 & $ -136.386\pm 1.865$ & $-137.343$ & $   0.957$ \\
2461079.35909 & 0.021 & $   32.757\pm 2.232$ & $  26.160$ & $   6.596$ \\
2461079.36738 & 0.063 & $  -48.731\pm 2.162$ & $ -41.697$ & $  -7.033$ \\
2461079.37567 & 0.105 & $ -103.590\pm 1.994$ & $-102.456$ & $  -1.133$ \\
\hline
\end{tabular}
\end{table*}

\begin{table*}
\centering
\caption{Parameters of the radial velocity curves
         $v_r(t)=\gamma+K_2 \sin[2\pi (t-T_0)/P]$ (see
         Figs~\ref{fig:ztfvel} and~\ref{fig:atlvel}). Values of $T_0$ and the
         period $P$ are from the time series photometry of the stars [see
         equations~(\ref{eq:eph_ztf}) and (\ref{eq:eph_atl})]. The last column
         contains the root mean square of the model residuals.}
\label{tab:rvcurve}
\begin{tabular}{crcccc}
\hline\hline
Star     & $\gamma$       & $K_2$          & $T_0$                 & $P$               & RMS \\
         & (km\,s$^{-1}$) & (km\,s$^{-1}$) & (HJD)                 & (d)               & (km\,s$^{-1}$) \\
\hline
ZTF~0329 & 136(22) & 391(64) & 2459831.6070(4.4E-4) & 0.1437540(2.3E-7) & 25 \\
ATL~0602 &  60(5)  & 266(4)  & 2459206.5515(2.9E-4) & 0.1964227(1.0E-7) & 12 \\
\hline
\end{tabular}
\end{table*}

\section{ZTF~0329 Discussion}
\label{sec:ztfdis}

As mentioned above, the M4--M5 spectral type of the RD component corresponds to
a temperature range of 3030--3200~K, whereas the SED fits give 3156--3200~K.
In particular, if the three outlying photometric measurements are excluded,
then $T_{\rm eff} \approx 3180$~K. As far as the WD is concerned, the high
temperatures in Table~\ref{tab:sed} appear incompatible with the SALT spectra,
and we adopt $T_{\rm eff}=7000$~K from the spectral fitting.

It was mentioned in \ref{sec:LCs} that there is evidence for non-thermal radiation
in ZTF~0329.
\citet{Krushinsky2020} found a similar effect in the PCEB GPX-TF16E-48. The
authors ascribe this to accretion onto the WD from a stellar wind emanating
from the RD. Further evidence for a non-thermal origin of some of the $R$-band
radiation in ZTF~0329 is visible in the SALT spectra in
Figs.~\ref{fig:saltspect1} and \ref{fig:saltspect2}. There is a clear excess of radiation in the region
6400--6620~\AA, which lies neatly within the mid-range of the Cousins $R$
filter (see e.g.\ \citealt{Rodrigo2024}).\footnote{\url{https://svo2.cab.inta-csic.es/theory/fps/index.php}}

Figs.~\ref{fig:saltspect1} and \ref{fig:saltspect2} can be compared with fig.~1 in \citet{VanRoestel2025},
which shows a broad hump of excess radiation in the spectrum of
ZTF~J005400.33+142931.4. The latter is also a short period (4.1~h) PCEB,
consisting of a WD and an M3.5 RD. The radiation excess centered on
$\sim 4860$~\AA\ is ascribed to electron cyclotron radiation induced by a
111~MG magnetic field associated with the WD.

The discrete line spectrum due to cyclotron radiation is given by
\begin{equation}
\lambda_n=\frac{1.0711 \times 10^{12}}{nB}\;, \;\;\;\;\; n=1,2,\cdots
\label{eq:cyclotron}
\end{equation}
\citep[e.g.][]{Liu2023}, where the magnetic field strength $B$ is in Gauss and
wavelength in~\AA. Possible low harmonic ($n \le 3$) cases which would give an
excess near 6500~\AA\ are
\begin{itemize}
\item[(i)]
$\lambda_1 \sim 6500$~\AA, $B \sim 165$~MG. In this instance
$\lambda_2 \sim 3250$~\AA\ and the rest of the cyclotron spectrum is in the UV.
\item[(ii)]
$\lambda_2 \sim 6500$~\AA, $B \sim 82$~MG. Then
$\lambda_1 \sim 13000$~\AA\ and $\lambda_3 \sim 4330$~\AA.
This can possibly be ruled out due to the lack of an excess in the 2MASS $J$
filter (Fig.~\ref{fig:sed0329}) and near 4330~\AA\ (see Figs.~\ref{fig:saltspect1}
and \ref{fig:saltspect2}).
\item[(iii)]
$\lambda_3 \sim 6500$~\AA, $B \sim 55$~MG. In this case
$\lambda_1 \sim 19500$~\AA\ and $\lambda_2 \sim 9750$~\AA.
The former wavelength lies between the ranges covered by 2MASS $H$ and $K_S$,
but the latter is within $z^\prime$ filter, and close to the red cutoff of
PanSTARRS $z$ at $\sim 9300$~\AA. It is therefore conceivable that the two
bright outliers in Fig.~\ref{fig:sed0329} are due to the second cyclotron harmonic.
\end{itemize}

We proceed to derive a rough estimate of the mass of the WD. From
Table~\ref{tab:rvcurve}, $K_2 \approx 391$~km\,s$^{-1}$ and it follows that
the mass function is
\begin{eqnarray}
f_M &=& \frac{M_1^3 \sin^3 i}{(M_1+M_2)^2}\nonumber\\
  &=& \frac{PK_2^3}{2\pi G} \nonumber\\
  &\approx& 0.890\,M_\odot \nonumber
\end{eqnarray}
i.e. $M_1 \ge 0.89\,M_\odot$.
A typical mass of an M5 dwarf is
$0.16\,M_\odot$,\footnote{\url{http://www.pas.rochester.edu/~emamajek/EEM_dwarf_UBVIJHK_colors_Teff.txt}}
hence $M_1\approx 1.18\,M_\odot$ ($i=90^\circ$) or
$M_1 \approx 1.28\,M_\odot$ ($i=75^\circ$).
For earlier RD spectral types the mass would be slightly higher (e.g.\
$0.22\,M_\odot$ for an M4 star), which would lead to slightly increased mass
estimates for the WD.

\section{ATL~0602 Discussion}
\label{sec:atldis}
\subsection{General}

Temperatures of the RD derived from photometry and spectroscopy are in very
good agreement, the range being 3364--3410~K. As for ZTF~0329 the photometric
temperature of the WD seems incompatible with the spectra, and we adopt
5750--6000~K for this component.

\begin{table}
\centering
\caption{Details of the {\it TESS} observations of ATL~0602. Photometry from
         sectors 6 and 87 are KSPSAP (optimal aperture size) or DET
         (detrended) magnitudes from the Quick-Look Pipeline
         \citep{Huang2020}, while sector~33 photometry are PDCSAP (pre-search
         data conditioned simple aperture photometry) magnitudes supplied by
         the Science Processing Operations Center \citep{Jenkins2016}.}
\label{tab:tess}
\begin{tabular}{ccccc}
\hline\hline
Sector & Interval covered & Time resolution & $N$   & Source \\
       & (TJD)            & (min)           &       &        \\
\hline
 6 & 1468--1490 &  30   &   989 & QLP  \\
33 & 2202--2228 &  10   &  3485 & SPOC \\
87 & 3664--3689 &  3.33 & 10535 & QLP  \\
\hline
\end{tabular}
\end{table}

There are two striking differences between the two LAMOST spectra
(Fig.~\ref{fig:lamost2}): emission lines are much stronger in LAMOST2, and it
shows a clear excess shortwards of $\sim 4600$~\AA\ when compared to LAMOST1.
The phase interval covered by LAMOST1 included an eclipse of the WD, which
most likely explains the lack of a blue excess in that spectrum.

Properties of the emission lines in the two spectra are summarised in
Table~\ref{tab:emlines}. These were obtained by fitting Gaussians to the line
profiles. Continuum levels were estimated by fitting a single straight line to
the spectra over intervals on either side of the line.

\begin{table*}
\centering
\caption{Properties of emission lines in the ATL~0602 spectra, obtained by
         fitting Gaussians to the line profiles. See the text for a description
         of the estimation of rotational broadening.}
\label{tab:emlines}
\begin{tabular}{ccccccccccc}
\hline\hline
  &  &  &  &\multicolumn{3}{c}{LAMOST1} && \multicolumn{3}{c}{LAMOST2}\\
Line & $\lambda_0$ & $\Delta \lambda_0$ && $V_r$ & FWHM & $v \sin i$ &&
$V_r$ & FWHM  & $v \sin i$  \\
     & (\AA) & (\AA) && (km\,s$^{-1}$) & (\AA) & (km\,s$^{-1}$) &&
(km\,s$^{-1}$) & (\AA) & (km\,s$^{-1}$) \\
\hline
H$\alpha$  & 6562.79 & 4.4 && $-92(3)$  & $6.4(0.2)$ & 122(21)  && 267(3)  & 6.4(0.1) & 122(10)  \\
H$\beta$   & 4861.35 & 3.2 && $-106(7)$ & $4.5(0.3)$ & 112(35)  && 264(2)  & 4.2(0.7) &  98(80)  \\
H$\gamma$  & 4340.47 & 2.9 && $-56(6)$  & $4.1(0.2)$ & 116(25)  && 256(5)  & 3.7(0.2) &  95(24)  \\
H$\delta$  & 4101.73 & 2.7 && $-52(24)$ & $5.7(0.8)$ & 187(126) && 245(9)  & 4.2(0.3) & 132(40)  \\
Ca~II~H    & 3968.47 & 2.7 && $-54(8)$  & $4.7(0.2)$ & 159(29)  && 299(10) & 5.1(0.3) & 173(46)  \\
Ca~II~K    & 3933.66 & 2.6 && $-78(13)$ & $5.5(0.4)$ & 189(64)  && 263(7)  & 4.3(0.2) & 144(28)  \\
\hline
\end{tabular}
\end{table*}

\begin{table*}
\centering
\caption{Attributes of the BM3 models fitted to the light curves of ATL~0602.
         Columns 5 and 6 contain the Roche lobe fillout factors of the two
         stars, and the final column is the RMS of the fits in
         Fig.~\ref{fig:bm3}. All spots are at $90^\circ$ colatitude, i.e.\ on
         the equator of the RD. The longitude is measured counter-clockwise
         from the line connecting the two stars. Spot temperatures are given as
         fractions of the RD temperature.}
\label{tab:bm3}
\begin{tabular}{ccccccccc}
\hline\hline
 & \multicolumn{8}{c}{Stellar properties}\\
Filter & $q$ & $R_1/A$ & $R_2/A$ && $f_1$ & $f_2$ && $\sigma$ \\
\hline
$I$ & 0.468 & 0.004  & 0.340 && $-0.988$ & $0.00081$ && 0.011 \\
$R$ & 0.451 & 0.015  & 0.340 && $-0.957$ & $0.034$   && 0.016 \\
$V$ & 0.453 & 0.013  & 0.336 && $-0.963$ & $-0.001$  && 0.015 \\
$B$ & 0.451 & 0.0135 & 0.338 && $-0.962$ & $0.012$   && 0.018 \\
\hline
 & \multicolumn{3}{c}{Spot properties} & & & & & \\
     & Longitude & Radius & Temperature  & & & & & \\
     & (deg)     & (deg)  &              & & & & & \\
\hline
$I$ & 155 & 13 & 0.85 & & & & & \\
$R$ & 155 & 13 & 0.76 & & & & & \\
$V$ & 180 & 12 & 0.80 & & & & & \\
$B$ & 180 & 14 & 0.80 & & & & & \\
\hline
\end{tabular}
\end{table*}

\begin{figure*}
\centering
\includegraphics[width=\textwidth]{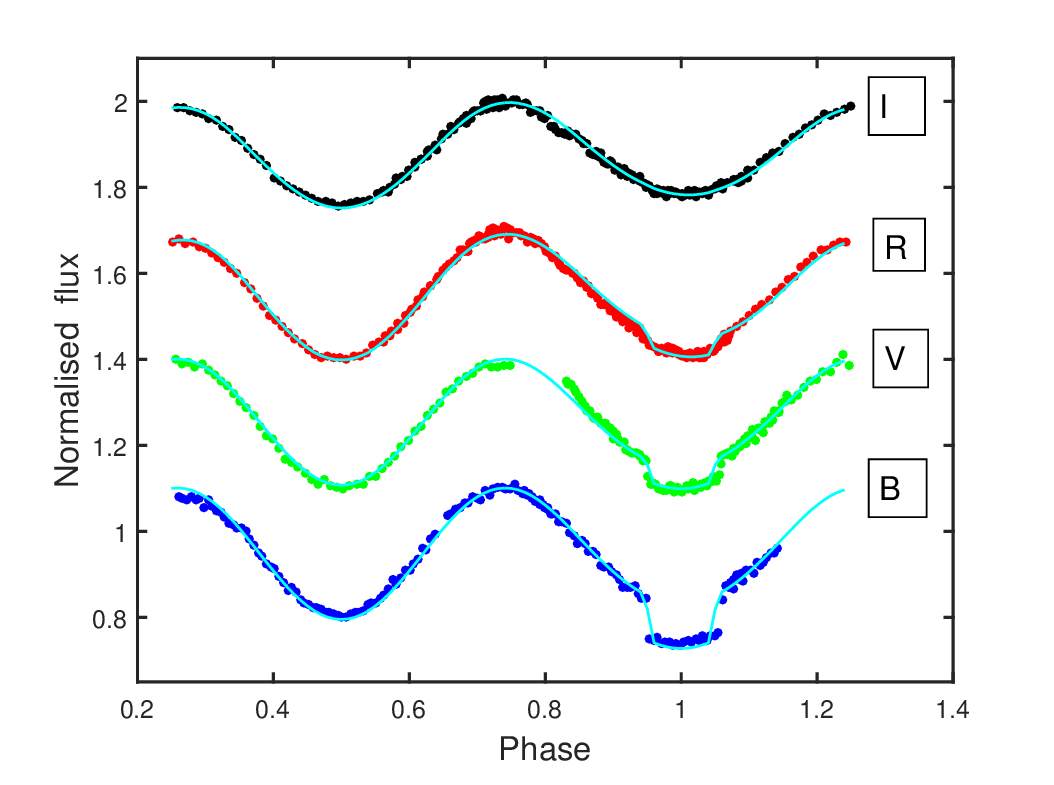}
\caption{Binary Maker model fits to the SAAO photometry of ATL~0602. The
         models include a cool spot on the surface of the RD component.}
\label{fig:bm3}
\end{figure*}

Values of $v\sin i$ in Table~\ref{tab:emlines} are calculated from the FWHMs
$\Delta \lambda$ of the spectral lines using an approximation in
\citet{Jackson2015}:
\begin{equation}
\Delta \lambda=\Delta \lambda_0 \left \{1+\left [R_\lambda
\frac{v \sin i} {K(u)c} \right ]^2 \right \}^{1/2}
\label{eq:fwhm}
\end{equation}
where $K$, a function of the linear limb-darkening coefficient $u$, is given by
$$K(u)= \left [ \frac{1-u/3}{2 \ln(2) (1-7u/15)} \right ]^{1/2}$$
with $R_\lambda$ the spectral resolution. In equation~(\ref{eq:fwhm}),
$\Delta \lambda=\lambda_0/R_\lambda$ is the spectral line width due to
instrumental broadening. It follows that
\begin{equation}
v\sin i \approx \frac{K(u) c}{R_\lambda} \sqrt{\frac{\Delta \lambda}
{\Delta \lambda_0}-1}
\label{eq:vsini}
\end{equation}

The Balmer and Ca~II line wavelengths in Table~\ref{tab:emlines} are covered by
various standard filters -- Johnson $U$ and $B$, Cousins $R_C$, Str\"omgren
$v$ and Sloan $g^\prime$ -- with limb darkening coefficients in
\citet{Claret2012}. Assuming $T_{\rm eff}=3400$~K and $\log\,g=5.5$,
$0.62 \le u \le 0.71$, and the constant $K(u)$ is in the interval
$[0.897, 0.907]$. For present purposes the assumption $K(u)=0.9$ is
sufficiently accurate, and hence
\begin{equation}
v\sin i \approx \frac{0.9\,c}{R_\lambda} \sqrt{\frac{\Delta \lambda}
{\Delta \lambda_0}-1}
\label{eq:vsini2}
\end{equation}

The differences between the velocities of the different lines are surprising.
Comparing the radial velocities in Tables~\ref{tab:specmod}
and~\ref{tab:emlines} it is evident that the emission lines are more closely
associated with the RD than the WD. If it is assumed that tidal forces has
synchronised rotation of the RD with the orbital period, then using the radius
from Table~\ref{tab:sed} it follows that $v_{\rm rot} \approx 148$~km\,s$^{-1}$.
The values of $v\sin i$ in Table~\ref{tab:emlines} are therefore consistent
with broadening by rotation of the RD. On the other hand, the Keplerian
velocity associated with a representative $0.7\,M_\odot$ WD with $\log\,g=9$
is $\sim 780$~km\,s$^{-1}$ even at a distance of 50 stellar radii from the WD,
making it very unlikely that the lines originate anywhere near the WD.

One possible explanation for the presence of the lines is that they are due to
irradiation by the WD component. There are two lines of evidence in support of
some degree of irradiation. First, we expect the irradiated face of the RD to
be most visible to the observer around phase 0.5. Given the phase intervals
covered by the two spectra, this would mean that emission lines should be
stronger in LAMOST2 -- as is observed. The second point is the radius inflation
of the RD. For reasonable masses of $M_1=0.8\,M_\odot$ and
$M_2=0.36\,M_\odot$ (see below), the length $A$ of the semi-major axis of the
binary orbit is
\begin{equation}
A =4.208\,P^{2/3}(M_1+M_2)^{1/3}=1.50\,R_\odot
\label{eq:sma}
\end{equation}
(where $P$ is the binary period in d). Given the value $R_2=0.57\,R_\odot$ in
Table~\ref{tab:sed}, the fraction of the WD radiation intercepted by the RD is
roughly
$$\frac{2\pi R_2^2}{4\pi (A-R_2)^2} \sim 0.19\;.$$

An argument against irradiation as the primary source of the emission lines is
the relatively small temperature difference ($\sim 2500$~K) between the two
stars. In all well-studied cases of irradiation of RDs by WDs in close binaries
the latter star appears to be hot, with temperatures of tens of thousands of
degrees \citep[e.g.][]{Catalan1995,Exter2005,Wawrzyn2009}.

It is also noted that the PCEB RR~Caeli is similar to 
ATL~0602, comprising a cool WD
($T_{eff} \sim 7540$ K) and an M4 RD \citep{maxted2007}. In that
system \citet{Bruch1999}
found that the H$\alpha$ emission strength was
strongest near primary eclipse, and concluded that this argues against
irradiation of the RD as the source of the emission. The
opposite is seen in the case of ATL~0602, but based on only
two spectra -- obviously not enough to draw firm conclusions about
the phase dependence of the emission.

Generally, the magnetic activity levels in stars increase with decreasing
rotation periods in late type stars. However, according to \citet{Freund2024}
almost all rapidly rotating stars with {\it Gaia} colour indices $B_p-R_p>2$
have saturated magnetic activity, i.e. the activity level does not increase
with increasing rotation. Given the very low level of reddening
(Table~\ref{tab:sed}), $(B_p-R_p)_0 \approx (B_p-R_p)=2.37$ for ATL~0602
\citep{GaiaCollab2023}, so that it is likely that the star is in the saturated
regime with $L_X/L_{\rm bol}\sim 10^{-3}$ \citep[e.g.][]{Jeffries2011}.
\citet{Freund2024} determined the X-ray flux of the star to be
$F_X=8.96$~mW\,m$^{-2}$; using its distance of 117~pc
\citep{BailerJones2021}, $L_X=3.83\times 10^{-5}\,L_\odot$. The bolometric
magnitude in Table~\ref{tab:sed} is easily converted to find
$L_{\rm bol}=0.0380\,L_\odot$, giving an observed value of
$L_X/L_{\rm bol}=1.01 \times 10^{-3}$, which confirms the saturated magnetic
activity. It follows that there is vigorous chromospheric activity in the star,
which could give rise to the emission lines.

Explanations for the variety of velocities of the lines are less obvious. The
simplest is that the wavelength shifts are due to calibration errors in the
LAMOST spectra. This topic has recently been studied by \citet{Zhang2026}, who
show velocity discrepancies up to 20--30~km\,s$^{-1}$ over the range
3900--4400~\AA\ (see their fig.~4). Generally, the velocity offsets are
negative for $\lambda<4900$~\AA, with the size of the velocity calibration
errors increasing with decreasing wavelength. As will be shown below, the
binary is marginally in a semi-detached configuration, with the RD close to
filling its Roche lobe. This suggests the possibility that lines originate, at
least in part, in mass escaping from the RD and flowing towards the WD.

From the radial velocity semi-amplitude $K_2 \approx 266$~km\,s$^{-1}$ and
period $P=0.1964$~d (Table~\ref{tab:rvcurve}) it follows that the mass function
is $f_M \approx 0.383\,M_\odot$. Eclipses of the WD are total, suggesting
an orbital inclination angle not too far from $90^\circ$. For an assumed mass
of $0.36\,M_\odot$ for an M3
dwarf\footnote{\url{http://www.pas.rochester.edu/~emamajek/EEM_dwarf_UBVIJHK_colors_Teff.txt}}
and $i=90^\circ$, $M_1 =0.80\,M_\odot$ (this changes to $0.83\,M_\odot$ for
$i=80^\circ$) and hence the mass ratio is $q \sim 0.44$.

The Eggleton zero-temperature mass-radius relation for WDs is
\begin{equation}
R_1=0.0114\,(M_1/M_C)^{-1/3} \left [ 1-(M_1/M_C)^{4/3} \right]^{1/2}
    \; R_\odot
\label{eq:eggleton}
\end{equation}
where $M_C=1.44$ is the Chandrasekhar WD mass limit \citep{Verbunt1988}.
Inspection of e.g.\ fig.~9 in \citet{Parsons2017} shows that in this mass
range radii of WDs with $T_{\rm eff}=10000$~K are very similar, so
equation~(\ref{eq:eggleton}) is sufficiently accurate. For the mass derived above, $R_1=0.0099\,R_\odot$. 
Exactly the same radius is obtained by assuming $T_1=6000$ K and
interpolating in the mass-radius table for DA WDs calculated by \citet{althaus2013}
and \citet{camisassa2016}.
\footnote{\url{https://evolgroup.fcaglp.unlp.edu.ar/modelos.html}}

Using the orbit size in
equation~(\ref{eq:sma}), $R_1/A \approx 0.0066$ and $R_2/A \approx 0.38$.

\subsection{Binary light curve modelling}

The latter values can be compared to those obtained by modelling of the binary.
Binary Maker 3
(BM3)\footnote{\url{http://www.binarymaker.com}} \citep{Bradstreet2002} was
used for that purpose. This software has the virtues of ease of use, and the
facility to switch between specification of stellar radii, Roche lobe filling
factors, and potentials. Both the O'Connell effect and slopes in the light
curves of the primary eclipses in Fig.~\ref{fig:atl0602lc} suggest the presence
of starspots, as does the changing light curve shapes in Fig.~\ref{fig:tessfig},
hence spots need to be included.

For the modelling, the temperatures of the two stars were kept fixed at
$T_1=6000$~K and $T_2=3400$~K, and the inclination was set at $i=90^\circ$.
Limb darkening coefficients were taken from \citet{Claret2012} and
\citet{Claret2020}. Models were fitted separately to the light curves obtained
through the different filters. The results are given in Table~\ref{tab:bm3}.

The fillout factor $-1<f \le 1$ is defined such that $f<0$ for stars
underfilling their inner Roche lobes (detached); $f=0$ when exactly filling the
inner critical surface (contact); $f>0$ for overfilling the inner Roche
(overcontact); and $f=1$ for stars filling their outer critical surfaces. The
small values of $f_1$ reflect the compact nature of the WD component, while
$f_2 \approx 0$ indicates that the RD is in marginal contact, i.e. the system
is in a semi-detached configuration.

There is good agreement between the mass ratios and stellar radii derived from
the $BVR$ light curves. The reason for the discrepant $I$ filter results is
that the primary eclipse is rounded in shape, rather than angular as for the
other three filters (see Fig.~\ref{fig:bm3}). This forces the solution towards
a smaller WD radius and larger value of $q$. It is noteworthy that the mass
ratios are in reasonable agreement with $q=0.44$ derived from the simple
estimate above, and $R_2/A$ is also similar to the value 0.38 estimated above.
On the other hand, $R_1/A$ in Table~\ref{tab:bm3} is more than double what is
expected on the basis of Eggleton's formula.

The light curve fits require cool spots of radius $\sim 13^\circ$ on the
equator of the RD. Note that different longitudes are indicated for different
filters -- the reason lies in the different slopes of the primary eclipses.
(Compare, for example, the $R$ and $B$ light curves in Fig.~\ref{fig:bm3}.)
The eclipses of the WD were more extensively observed in $R$ and $I$ in 2020,
while the $B$ and $V$ observations were restricted to 2025, which could explain
the differences in the respective best-fitting spot configurations found. If
this is correct, the agreements in properties other than the longitudes (spot
sizes and temperatures) across filters is remarkable.

\section{Conclusions}
\label{sec:conc}

There are a number of questions raised by the observations reported in this
paper. One of these is the close resemblance between the independently obtained
outlying PanSTARRS $z$ and SDSS $z^\prime$ measurements of ZTF~0329, while the
SkyMapper $z$ is consistent with the rest of the photometry. Of course, the
changes in light curve shape (Fig.~\ref{fig:ztf0329lc}) and in the mean
brightness level, might mean that some measurements are highly epoch-dependent.

If the excess $R$ band radiation in ZTF~0329 is indeed due to electron
cyclotron radiation, then a near infrared spectrum of the star could clarify
the harmonic order. This would then also establish the strength of the magnetic
field associated with the WD. Of course, polarisation measurement may also
reveal properties of the magnetic field. Study of polarisation in ATL~0602
spectral lines could help discriminate between irradiation and chromospheric
activity as primary drivers of the emission.

It has been proposed that mass motions in the binary system may be responsible
for the O'Connell effect in contact binaries
\citep[e.g.][]{Zhou1990,Fabry2025}. This is an attractive idea in the present
context of the semi-detached ATL~0602, as it could also explain the peculiar
array of emission line radial velocities found in that system. The reader is
also referred to the discussion by \citet{Tappert2011} of multiple-component
H$\alpha$ emission lines found in a number of PCEBs.

\section*{Data Availability}

SAAO photometry is available from the first author, and the SALT spectra from
the second author. Other data can be obtained from sources given in the text.

\section*{Acknowledgements}

Allocation of telescope time by the South African Astronomical Observatory and
the Southern African Large Telescope is acknowledged. The authors are
particularly grateful for Director's Discretionary Time awarded in order to
obtain SALT spectra of ZTF~0329. This research has made use of the VizieR
catalogue access tool and the Simbad Astronomical Database at CDS, Strasbourg,
France; the collection of stellar model spectra and filter transmission
functions of the Spanish Virtual
Observatory;\footnote{\url{http://svo2.cab.inta-csic.es/theory/main/}} bolometric
corrections from the ``MESA Isochrones and Stellar
Tracks'';\footnote{\url{http://waps.cfa.harvard.edu/MIST/model_grids.html\#bolometric/}}
and the results of various large photometric surveys referred to in
Sections~\ref{sec:sed} and~\ref{sec:specmod} of the paper.
A.K. acknowledges support from the National Research Foundation (NRF) of South
Africa.

\bibliographystyle{mnras}
\bibliography{WD2_mnras}

\bsp
\label{lastpage}

\end{document}